\documentclass[12pt, draftclsnofoot, onecolumn]{IEEEtran}
\pdfoutput=1

\IEEEoverridecommandlockouts

\normalsize
\usepackage{cite}
\usepackage[normalem]{ulem}
\usepackage{color}
\usepackage{graphicx} 
 \ifCLASSOPTIONcompsoc
	\usepackage[caption=false,font=footnotesize,labelfont=sf,textfont=sf]{subfig}
\else
	\usepackage[caption=false,font=footnotesize]{subfig}
\fi
\RequirePackage{lineno}

\usepackage{color}
\usepackage[usenames]{xcolor}

\usepackage{arydshln,leftidx,mathtools}
\usepackage{booktabs}
\usepackage[inline]{enumitem}
\usepackage{fixltx2e}

\usepackage{amsmath,amssymb,amsfonts,bbm}
\usepackage{arydshln,leftidx,mathtools}
\usepackage[backgroundcolor=cyan,bordercolor=gray,textsize=small,textwidth=5.8cm]{todonotes}

\usepackage{url}

\usepackage{algorithm}
\usepackage{algpseudocode}
\usepackage[flushleft]{threeparttable}

\DeclareMathOperator*{\rank}{rank}

\usepackage{algorithm}
\usepackage{algpseudocode}
\usepackage[flushleft]{threeparttable}

\makeatletter
\def\BState{\State\hskip-\ALG@thistlm}
\makeatother

\usepackage{textcomp}

\newtheorem{proposition}{Proposition}

\newcommand{\blackcircle}{\tikz[baseline]{\draw[black,solid,line width = 0.7pt](2.7mm, 1mm) circle (0.8mm)}}
\newcommand{\bluesquare}{\tikz[baseline]{\draw[blue,solid,line width = 0.7pt] (0.2mm,0.2mm) -- (1.8mm,0.2mm) -- (1.8mm,1.8mm) -- (0.2mm,1.8mm) -- (0.2mm,0.2mm)}}
\newcommand{\redcross}{\tikz[baseline]{\draw[red,solid,line width = 0.7pt](0.2mm,0.2mm) -- (1.8mm,1.8mm) -- (1mm,1mm) -- (1.8mm,0.2mm) -- (0.2mm,1.8mm)}}
\newcommand{\magentadiamond}{\tikz[baseline]{\draw[magenta,solid,line width = 0.7pt](0mm,1mm) -- (0.7mm,2mm) -- (1.4mm,1mm) -- (0.7mm,0mm) -- (0mm, 1mm)}}

\newcommand{\grayplus}{\tikz[baseline]{\draw[gray,solid,line width = 0.7pt](0mm,1mm) -- (1.8mm,1mm) -- (0.9mm,1mm) -- (0.9mm,1.8mm) -- (0.9mm,0.2mm)}}
\newcommand{\greentriangle}{\tikz[baseline]{\draw[green,solid,line width = 0.7pt](0.2mm,1.8mm) -- (1.8mm,1.8mm) -- (1mm,0.2mm) -- (0.2mm,1.8mm) -- (1.8mm,1.8mm)}}

\newcommand{\blackline}{\tikz[baseline]{\draw[black,solid,line width = 0.9pt](0mm,1mm) -- (5.0mm,1mm)}}
\newcommand{\magentaline}{\tikz[baseline]{\draw[magenta,solid,line width = 0.9pt](0mm,1mm) -- (5.0mm,1mm)}}
\newcommand{\blueline}{\tikz[baseline]{\draw[blue,solid,line width = 0.9pt](0mm,1mm) -- (5.0mm,1mm)}}
\newcommand{\redline}{\tikz[baseline]{\draw[red,solid,line width = 0.9pt](0mm,1mm) -- (5.0mm,1mm)}}

\begin{document}

\title{Optimum Multi-Antenna Ambient Backscatter Receiver for General Binary-Modulated Signal}

\author{Xiyu Wang,
H{\"u}seyin~Yi\u{g}itler and Riku~J\"{a}ntti
\thanks{
The authors are with the Department of Communications and Networking, Aalto University, Espoo, 02150 Finland.
(e-mail: firstname.surname@aalto.fi)
}
}

\maketitle

\begin{abstract}
Ambient backscatter communication (AmBC) is becoming increasingly popular for enabling green communication amidst the continual development of the Internet-of-things paradigm. Efforts have been put into backscatter signal detection as the detection performance is limited by the low signal-to-interference-plus-noise ratio (SINR) of the signal at the receiver. The low SINR can be improved by adopting a multi-antenna receiver. In this paper, the optimum multi-antenna receiver that does not impose any constraints on the types of binary modulation performed by the backscatter device and the waveform used by the ambient source system is studied. The proposed receiver owns a simple structure formed by two beamformers. Bit error rate (BER) performances of the optimum receiver are derived under constant-power ambient signal and Gaussian-distributed ambient signal. Moreover, to facilitate the implementation of the optimum receiver, a simplified receiver is proposed and practical approximations to required beamformers are provided.  The derived optimum receiver avoids the complex direct path interference cancellation and coherent reception, but exploits the fact that backscatter signal changes the composite channel impinging at the receiver and the directivity of receiver antenna array. 
Comparative simulation results show that the performance of the optimum receiver achieves the same performance as the coherent receiver even though it realizes non-coherent reception. The studied receivers provide high flexibility for implementing simple and low-cost receivers in different AmBC systems.

\end{abstract}  

\section{Introduction}
%general -- problem -- SOA -- challenge -- our contribution
% previous researches make assumption on the ambient source signal

Internet-of-Things (IoT) technologies continue to be rolled out around the world, consisting of wide-area use cases.  The number of IoT connections is approximately incremented by a factor of 3 during the past year and is forecast to reach 25 billion by 2025 ~\cite{Ericsson2019}.    The increasing deployment density of sensors inevitably consumes a great amount of both energy resource and spectrum resources, which in turn limits IoT networks. The recent emerging ambient backscatter communication (AmBC) is a solution relieving both of these two limitations. 	In a typical AmBC deployment scenario (see Fig.~\ref{fig:systemModel}), a transmitter of legacy system emits the radio frequency (RF) signal serving its legacy devices. A passive backscatter device (BD) harvests energy from the pervasive ambient RF signal to support its operations and transmits its own information by modulating on top of the RF signal. A composite of signal paths, the \emph{backscatter path} which is modulated by the BD, and the \emph{direct path} which is not affected by BD operations, impinge at the receiver, from which the receiver decodes the backscatter signal.  Requiring neither power-hungry nor expensive radio frequency (RF) components for the sensor circuits, AmBC paradigm realizes the ultra-low power and ultra-low cost green communication ~\cite{Liu2013}. It further provides significant bandwidth efficiency by enabling data exchange among simple devices without a dedicated reader generating specific carrier signal for sensors which occupies spectrum resource but using the spectrum allocated for a legacy system.  With these two inherent properties, AmBC is promising to become an important component for realizing sustainable IoT communication.

The widespread acceptance of AmBC system is limited by the poor detection performance for the desired backscatter signal. One factor that hampers the detection performance is the strong direct path interference (DPI) resulting in a tremendous power degradation of the concatenate backscatter channels.  Moreover,  backscatter systems lack cooperation with legacy systems so that AmBC receivers have little information about ambient RF signal.% which causes fast fading to the backscatter signal. 
Owing to the above two factors, the backscatter path experiences a low signal-to-interference-plus-noise ratio (SINR) which limits the detection performance of AmBC system.

Existing approaches that boost detection performance via addressing DPI and unknown ambient signal include exploiting frequency~\cite{Yang2018OFDM, ElMossallamy2019}, spatial~\cite{Ma2015, Huayan2019, Duan2019, Tao2020JIoT} or  phase~\cite{Anyscatter2020} differences between two paths, and using complex signal processing techniques \cite{Yang2018Cooperative}. Among these solutions, receiver with multiple antennas that exploits spatial difference has attracted much attention in AmBC researches. It is able to mitigate strong DPI and provide partial estimate of ambient RF signal while requiring neither further assistance from legacy system nor extra information about channels. Therefore, adopting multiple antennas at the receiver is a desire and practical method of improving detection performance. 
	
Available works on multi-antenna AmBC receivers mostly consider systems where the BD performs on-off-keying (OOK) modulation. An optimum multi-antenna receiver for OOK-modulated backscatter signal is derived in \cite{Tao2020JIoT}. Though OOK is the most commonly adopted modulation on a BD, its demodulator loses a certain SNR gain with respect to bit-error-rate (BER)-performance compared with other binary demodulators. Recovering backscatter signal of different modulations, typically, needs different receiver structures. Such fact causes the implementation of AmBC systems lacks flexibility and has uncertain computational complexity. To take binary phase shift keying (BPSK) as an example, its demodulator can achieve the same BER performance with up to 6 dB less SNR compared with OOK demodulator, and its optimum multi-antenna receiver obtaining all the SNR gain is a coherent receiver which is analyzed in work \cite{Xiyu2020CoherentReceiver}. However, the optimum multi-antenna receivers for other binary modulations have not been studied.

In this paper, we investigate the optimum multi-antenna receiver that works for any binary-modulated backscatter signal.  Derived from the maximum-a-posteriori (MAP) criterion, we obtain the optimum multi-antenna receiver for AmBC system, based on which we propose a simplified version of the optimum receiver to facilitate the practical implementation. Then, we analyze the performance of the optimum receiver under two types of ambient signal: unknown signal with constant amplitude and Gaussian-distributed signal.
We analyze the simplified receiver using \emph{Neyman-Pearson} criterion. We provide and compare three practical methods for estimating the required beamformers.

Our derivation results show that the optimum receiver has a very simple and clear structure. It avoids the complex DPI cancellation and coherent reception, but explores the directivity of the composite channel impinging at the receiver. The optimum receiver is effective when the direction of the composite channel changes as backscatter signal changes. The optimum receiver has no constraint on the type of modulation of neither backscatter signal nor ambient signal.
The primary contributions of this paper are as follows.

\begin{itemize}
\item We formulate and solve the optimum multi-antenna receiver for general binary modulated backscatter signal in AmBC. The optimum receiver takes a form of energy detector and owns a very simple structure that contains two beamformers and the decision threshold of test statistic 0. The optimum receiver achieves the same BER-performance as the coherent reception of the BPSK-modulated backscatter signal, although it avoids demodulating signal coherently.

\item We derive the error probabilities of the optimum receiver for two types of ambient signal: i) unknown ambient signal with constant amplitude and ii) Gaussian-distributed ambient signal with varying energy. The results suggest that AmBC systems under ambient signal with constant amplitude achieve the same BER-performance of AmBC systems under Gaussian-distributed ambient signal with at least 4-dB less SINR.

\item We propose a simplified receiver with only one beamformer to facilitate the practical implementation of the optimum receiver. Since the statistic information about backscatter signal is not available at the receiver, we analyze the simplified receiver based on \emph{Neyman-Pearson} criterion and give closed-form expressions for the probability of detection, false alarm and its ROC curve.

\item We also provide three practical and easy to implement methods for obtaining the beamformers that construct the test statistics of two studied receivers. Results show that taking the left singular vector associated with the largest singular value of sample matrix and power iteration are efficient methods giving close estimates to the ideal beamformers with a small length of preambles.

    % \item Finally, we provide simulation results to corroborate our analysis and to study the effects of various system parameters, namely, number of receive antennas, modulations of the BD and the source signal, spatial deployment of BD, transmitter and receiver, number of training samples. 
\end{itemize}

The remainder of the paper is organized as follows. Related works are reviewed in Section~\ref{sec:relatedWork}.
The system model is described in Section~\ref{sec:systemModel}. In Section~\ref{sec:ORformulation}, the optimum receiver is derived, and its performance is analyzed under two cases of ambient signal. The practical implementations for the optimum receiver including the simplified receiver and estimation methods are studied in Section~\ref{sec:pracImplementation}. Simulation results are provided in Section~\ref{sec:simulation} to evaluate our analysis. Finally, conclusions are drawn in Section~\ref{sec:conclusion}.

\section{Background}\label{sec:relatedWork}

\subsection{Related Work}
The aim of this paper is to investigate an optimum multi-antenna receiver for demodulating any binary-modulated backscatter signal. 
We first give a literature review of receivers designed for different binary modulations of backscatter signal.
Then, we review existing AmBC systems with a multi-antenna receiver.

Efforts have been devoted to improve the detection performance of AmBC system through mitigating the negative impacts of the strong DPI and unknown ambient signal. 
Available solutions for addressing these two factors are designed based on a specific setup of the AmBC system.
%Building upon the fact that backscatter devices transmit their information by modifying either amplitude, or phase, or frequency of the incoming ambient signal and then reflecting it to receivers~\cite{Xu2018Mag}. 
For backscatter device adopting the most commonly OOK modulation or differential BPSK modulation, the received signal strength varies as BD switches the signal state. Based on this, non-coherent receivers are implemented by energy detector and maximum likelihood receiver to average out the fast varying phases and compare the energy levels between two signal states~\cite{Parks2014,Liu2013,Wang2016,Qian2019,Zhang2019,Yang2018Cooperative,Yang2018OFDM, Huayan2019,Tao2020JIoT,Hwang2020, Idrees2020}. In these systems, 
DPI is mitigated by considering the AmBC system working under orthogonal frequency division multiplexing (OFDM) ambient signal and BD shifts the frequency of ambient signal~\cite{ElMossallamy2019, Darsena2019}, or by designing specific training sequence~\cite{Idrees2020}. Furthermore, the non-coherent receiver loses SNR gain compared with a coherent receiver.  Coherent receiver of AmBC system requires complex phase synchronization methods~\cite{Vougioukas2019} or additional cooperation between legacy system and backscatter system~\cite{Yang2018Cooperative} because phases of both ambient signal and channels are not known to the receiver. Different from the above proposed receivers, we propose a more general receiver structure that has no constraint on modulation of backscatter signal and type of ambient signal.

Adopting multiple antennas at AmBC receiver is gaining interest.
Prior works utilize the diversity provided by multiple antennas to detect backscatter signal. Multi-antenna receivers differentiate states of backscatter signal by investigating the difference of signal strength~\cite{Parks2014,Kang2017a} or difference of phase~\cite{Anyscatter2020} between antenna elements. However, these methods take advantage of the diversity of received signal among different antennas but lose the array gain of received signal. In contrast, our proposed receiver obtains array gain in addition to the diversity gain. 

Recent works also exploit multiple antennas at the receiver to alleviate the adverse impact of both DPI and unknown ambient signal. 
Specifically, a multi-antenna receiver mitigates the strong DPI through separating the desired backscatter path from the DPI using beamforming technique without additional information about ambient signal or channels~\cite{Huayan2019,Duan2019}. The work~\cite{Huayan2019} tests the energy of beamformed received signal to detect the OOK-modulated backscatter signal.  Additionally, prior work \cite{Tao2020JIoT} devises an optimum multi-antenna receiver for OOK-modulated backscatter signal and complex Gaussian ambient signal. On the contrary, we introduce an optimum receiver of AmBC system for general binary-modulated backscatter signal and different types of ambient signal. Our proposed optimum receiver owns a significantly simpler structure and the decision threshold is independent of signal and channels.

\subsection{Notations}
Throughout the paper, scalars are denoted by normal font letters $a$, vectors and matrices are represented by lower-case $\boldsymbol{a}$ and upper-case $\boldsymbol{A}$ boldface letters, respectively. Complex numbers are assumed and its set is denoted by $\mathbb{C}$. The Euclidean norm of vector $\boldsymbol{a}$ is represented by $\|\boldsymbol{a}\|$ and the absolute value of $a$ is represented by $|a|$.
The $n\times n$ identity matrix is $\boldsymbol{I}_n$, and the subscript $n$ may be omitted sometimes for simplicity.
The conjugate-transpose, conjugate, and transpose of matrix $\boldsymbol{A}$ are denoted as $\boldsymbol{A}^H$, $\boldsymbol{A}^*$ and $\boldsymbol{A}^T$, respectively.
The trace and the rank of matrix $\boldsymbol{A}$ are represented by $\mbox{tr}\{\boldsymbol{A}\}$ and $\rank (\boldsymbol{A})$.
We use $\mathcal{CN}(\boldsymbol{a}, \boldsymbol{A})$ to denote the circularly symmetric complex Gaussian variable with mean $\boldsymbol{a}$ and covariance matrix $\boldsymbol{A}$.  The statistical expectation is $\mathbb{E}\{\cdot\}$. The imaginary unit is $j=\sqrt{-1}$.

\section{System Model}\label{sec:systemModel}

\begin{figure}[!t]
    \centering
    \includegraphics[width=0.465\textwidth]{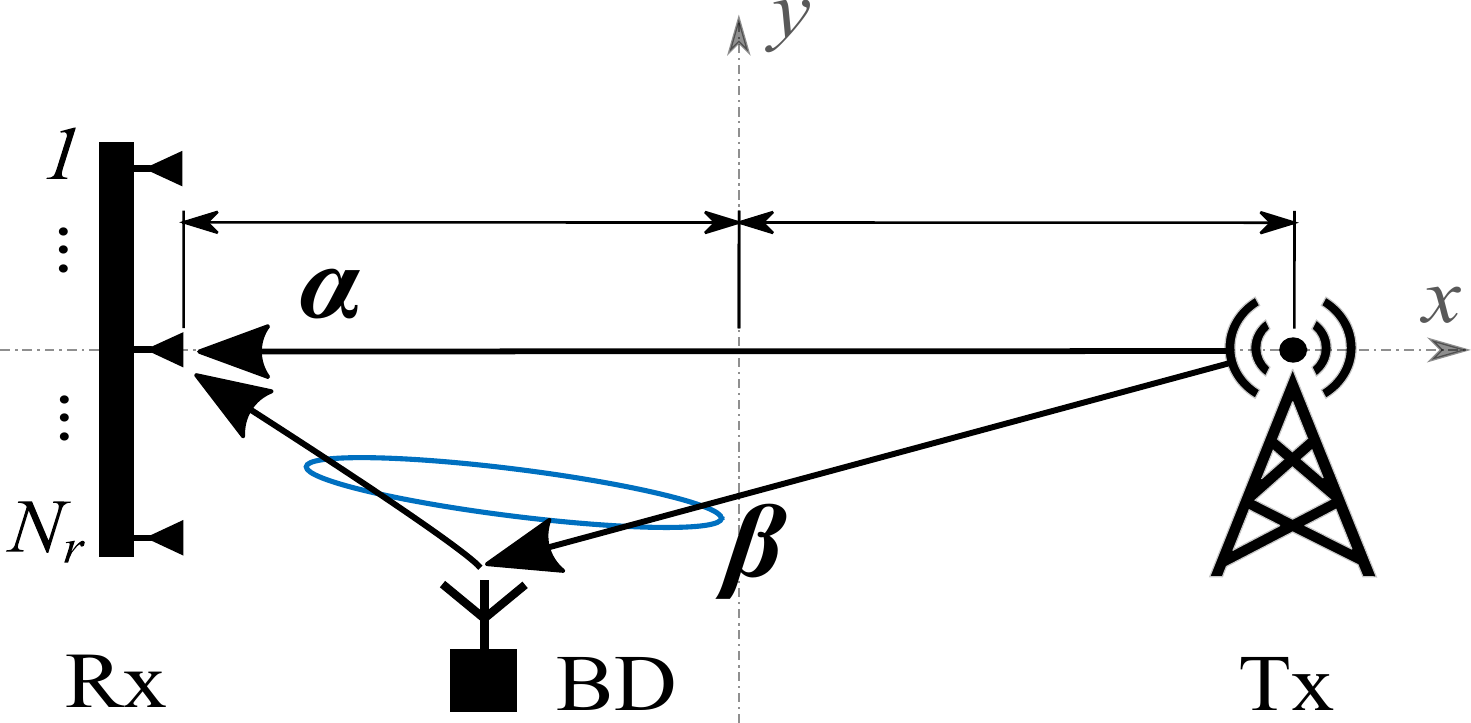}
    \caption{An illustration of the AmBC systems.}
     \label{fig:systemModel}
\end{figure}

In this paper, a typical bi-static narrow-band AmBC system, depicted in Fig.~\ref{fig:systemModel}, which consists of a legacy ambient source (Tx), a single-antenna backscatter device (BD) and an $N_r$-antenna AmBC receiver (Rx) is considered. Three nodes are placed in the Cartesian reference frame of a two-dimensional Euclidean space, as shown in the figure. The $ \lceil N_r/2 \rceil$-th antenna out of $N_r$ antennas on the Rx is selected to be the reference antenna. The line segment connecting the Tx antenna and the Rx reference antenna is set as x-axis while their middle point is set to be the origin of the reference frame. The locations of the Tx, the BD and the $\ell$-th, $\ell = 1,\cdots, N_r$, Rx antenna are denoted by $\boldsymbol{p}_t$, $\boldsymbol{p}$ and $\boldsymbol{p}_{r\ell}$, respectively. With locations of three nodes, the distances between the Tx and $r$-th Rx antenna and between the BD and $r$-th Rx antenna can be easily computed as
\begin{equation*}
        d_{0 \ell} = \|\boldsymbol{p}_t - \boldsymbol{p}_{\ell}\| ,~~
        d_{1 \ell} = \|\boldsymbol{p} - \boldsymbol{p}_{\ell}\|,~~ \mathrm{and} ~~
        d_{2} = \|\boldsymbol{p}_t - \boldsymbol{p}\|  .
\end{equation*}
For simplicity, the distances between Tx-Rx and BD-Rx are referred to the distances from the Rx reference antenna to the Tx and the BD, denoted by $d_0$ and $d_1$, respectively.

For narrow-band systems that cannot resolve individual multipath components, let $\boldsymbol{\alpha} = [\alpha_1,\cdots, \alpha_{N_r}]^T$ compose all multipath components impinge on the receiver antennas that are not modulated by the BD\footnote{This and other vectors are constant for block fading channels with coherence time exceeding several BD symbol times. Our following analysis is conducted within one channel coherent time.} which is termed as \emph{direct path}, and $\boldsymbol{\beta} = [\beta_1,\cdots,\beta_{N_r}]^T$ compose the multipath components that are modulated by the BD\footnote{Define $\Delta = \|\boldsymbol{\alpha}\|^2/\|\boldsymbol{\beta}\|^2$ as the power difference between the direct path and the backscatter path. It has been discussed in~\cite{Xiyu2019} thoroughly that $\Delta$ can reach up to 40 dB when the BD is 5 meters away from the Rx. Here we normalize the $\|\alpha\|^2 =1$.} which is termed as \emph{backscattered path}.
In particular, by applying a simplified Friis pathloss formula, the channel gain of the direct path of the $\ell$-th Rx antenna is 
\begin{equation*}
        \alpha_{\ell} = \sqrt{\left(\frac{\lambda}{4\pi d_{0\ell}}\right)^2}\exp\left\{j2\pi\frac{d_{0\ell}}{\lambda}\right\} ,
\end{equation*}
and the channel gain of the backscatter path of the $\ell$-th Rx antenna is
\begin{equation*}
\begin{aligned}
    \beta_{\ell} = \sqrt{\left(\frac{\lambda}{4\pi d_{2} }\right)^2 \left(\frac{\lambda}{4\pi d_{1\ell}}\right)^2 }\exp\left\{j2\pi\frac{d_2 + d_{1\ell}}{\lambda}\right\}  ,
\end{aligned}
\end{equation*}
where $\lambda = f_c/c_0$ is the carrier wavelength, $f_c$ is the carrier frequency and $c_0$ is the free-space electromagnetic propagation speed.

Let us denote the $k$-th binary-modulated backscatter signal transmitted from the BD as $x[k] \in \mathcal{X}=\{x_0^{},x_1^{}\}$ and 
the $k$-th unknown ambient signal from the Tx as $s[k]$.
The sample output of $N_r$ receiver antennas can be written as 
\begin{equation}\label{eq:received-signal:y}
    \boldsymbol{y}[k] =
        \boldsymbol{\alpha} s[k] + \boldsymbol{\beta} s[k] x[k] + \boldsymbol{n}[k],
\end{equation}
where $\boldsymbol{n}$ is zero-mean, identically distributed, standard circularly-symmetric complex Gaussian 
noise, i.e., $\boldsymbol{n}[k] \sim \mathcal{CN}(\boldsymbol{0},\boldsymbol{I}_{N_r})$, whose components are assumed to be independent of ambient signal and backscatter signal. Since noise $\boldsymbol{n}$ has normalized variance, the received signal-to-noise ratio (SNR) seen by the reference antenna is denoted by 
\begin{equation*}
    \gamma = \left(\frac{\lambda}{4\pi d_0}\right)^2|s|^2 .
\end{equation*}

In this paper, we assume BD adopts the binary modulation with equal probability, i.e., $p(x_0) = p(x_1)$. 
We further assume the Rx has no prior statistical information about the source signal. In the remaining part of the paper, we drop the time dependence of $\boldsymbol{y}$ since the scope of the analysis is restricted on decoding $x$ based on a single temporal sample of received signal $\boldsymbol{y}$. For later utilization, we define the compound channel gains of the direct path and the backscatter path as
\begin{equation*}
    \boldsymbol{g}(x) \triangleq \boldsymbol{\alpha} + x \boldsymbol{\beta}  .
\end{equation*}

The goal of this paper is to investigate an optimum multi-antenna receiver for general binary-modulated backscatter signal in AmBC system.

\section{Optimum Multi-antenna Receiver}\label{sec:ORformulation}
In this section, the derivation of the optimum multi-antenna receiver for AmBC system is elaborated on, followed by its performance analysis for two types of ambient signal, namely the signal with constant-power and Gaussian-distributed signal.

\subsection{Optimum Receiver Formulation}
In formulating the optimum receiver of multi-antenna AmBC system, we consider the maximum-a-posteriori probability (MAP) principle which makes a decision maximizing the probability of the backscatter signal given the received signal~\cite{Wozencraft1965}. 
Then, detecting the binary-modulated BD signal can be written as a binary hypothesis testing, that is
\begin{equation}\label{eq:MAPcriterion}
    p(x_0 |\boldsymbol{y}) \mathrel{\substack{\mathcal{H}_0\\ \gtrless \\ \mathcal{H}_1}} p(x_1|\boldsymbol{y}) .
\end{equation}

In AmBC system, the ambient source signal $s$ is a latent variable which is not straightforwardly available at the receiver due to the lack of cooperation between legacy system and AmBC system. If there exists a distribution for $s$, its effect on the posterior probability in Eq.~\eqref{eq:MAPcriterion} can be marginalized out, i.e.,
\begin{equation} \label{eq:posterioriProb}
    \begin{aligned} 
    p(x|\boldsymbol{y})&= \int_\mathbb{S} p(x,s|\boldsymbol{y}) ds = \int_\mathbb{S} p(x|\boldsymbol{y}, s) f(s|\boldsymbol{y}) ds  \\
    &= \int_\mathbb{S} \frac{f(\boldsymbol{y}|x,s) p(s|x)p(x)}{f(\boldsymbol{y},s)}\frac{f(\boldsymbol{y},s)}{f(\boldsymbol{y})} ds \\
    &= \int_\mathbb{S} \frac{f(\boldsymbol{y}|x,s) p(s)p(x)}{f(\boldsymbol{y})} ds
    \end{aligned} ,
\end{equation}
for $\forall x \in \mathcal{X}$, and $\mathbb{S}$ denotes the ambient signal space. 

We consider a worse yet practical case that prior statistical information about $s$ is not available at the receiver. Accordingly, we resort to the multiple antennas of the receiver to obtain the coarse estimate of the ambient signal $\hat{s}$. 
When the estimate of ambient signal is given, the probability $p(s) = \delta(s - \hat{s})$, where $\delta(\cdot)$ is the Dirac delta function. Then, Eq.~\eqref{eq:posterioriProb} is
\begin{equation*}
    \begin{aligned}
    p(x|\boldsymbol{y}) &= \int_\mathbb{S} \frac{ f(\boldsymbol{y}|x,s) \delta(s-\hat{s}) p(x)}{f(\boldsymbol{y})} ds \\
    &= \frac{f(\boldsymbol{y}|x,s=\hat{s}) p(x)}{f(\boldsymbol{y})} .
    \end{aligned}
\end{equation*}
Substituting it into Eq.~\eqref{eq:MAPcriterion} and taking logarithm on both sides yields the receiver based on log-likelihood criterion, written as
\begin{equation}\label{eq:LogMAPcriterionApprox}
   \ln f(\boldsymbol{y}|x_0^{},s=\hat{s}) \mathrel{\substack{\mathcal{H}_0\\ \gtrless \\ \mathcal{H}_1}} \ln f(\boldsymbol{y}|x_1^{},s=\hat{s}) ,
\end{equation}
as $x \in \mathcal{X}$ has equal probability, and  $f(\boldsymbol{y})$ is independent of $x$. 

\begin{proposition}
The log-likelihood function in Eq.~\eqref{eq:LogMAPcriterionApprox} is given by
\begin{equation} \label{eq:loglikelihood}
    \begin{aligned}
        &\ln{f(\boldsymbol{y}|x,s=\hat{s}^{\scalebox{0.7}{ML}})} \\=& - \boldsymbol{y}^H \boldsymbol{G}(x) \boldsymbol{y} - N_r \ln \pi .
    \end{aligned}
\end{equation}
\end{proposition}
\begin{IEEEproof}
See Appendix.~\ref{appendix:loglikelihood}.
\end{IEEEproof}

% Taking logarithm of the PDF gives 
% \begin{equation} \label{eq:loglikelihood}
%     \begin{aligned}
%         &\ln{f(\boldsymbol{y}|x,s=\hat{s}^{\scalebox{0.7}{ML}})} \\=& - \boldsymbol{y}^H \boldsymbol{G}(x) \boldsymbol{y} - N_r \ln \pi .
%     \end{aligned}
% \end{equation}
Substituting Eq.~\eqref{eq:loglikelihood} into Eq.~\eqref{eq:LogMAPcriterionApprox}, the binary hypotheses testing is simplified to 
\begin{subequations} 
\begin{gather}
    \boldsymbol{y}^H \boldsymbol{G}(x_0)  \boldsymbol{y}           \mathrel{\substack{\mathcal{H}_0\\ \lessgtr \\ \mathcal{H}_1}} 
   \boldsymbol{y}^H \boldsymbol{G}(x_1)\boldsymbol{y} \\
    \label{eq:energyComparison}    \implies ~~
   \boldsymbol{y}^H  \left(\boldsymbol{I} - \frac{\boldsymbol{g}_0^{} \boldsymbol{g}_0^H }{\|\boldsymbol{g}_0^{}\|^2}\right)  \boldsymbol{y}  \mathrel{\substack{\mathcal{H}_0\\ \lessgtr \\ \mathcal{H}_1}} 
  \boldsymbol{y}^H \left( \boldsymbol{I} - \frac{ \boldsymbol{g}_1^{} \boldsymbol{g}_1^H }{\|\boldsymbol{g}_1^{}\|^2} \right)\boldsymbol{y}
 \\
  \label{eq:testStatistic}      \implies ~~ z\triangleq
  \boldsymbol{y}^H  \left[\left( \boldsymbol{I} - \frac{ \boldsymbol{g}_1^{} \boldsymbol{g}_1^H }{\|\boldsymbol{g}_1^{}\|^2} \right)   -  \left(\boldsymbol{I} - \frac{\boldsymbol{g}_0^{} \boldsymbol{g}_0^H }{\|\boldsymbol{g}_0^{}\|^2}\right) \right]\boldsymbol{y} \mathrel{\substack{\mathcal{H}_0\\ \gtrless \\ \mathcal{H}_1}} 0 .
   \end{gather} 
\end{subequations}
where, for notational convenience, we denote 
\begin{equation*}
\boldsymbol{g}_0 = \boldsymbol{g}(x_0), \qquad
\boldsymbol{g}_1 = \boldsymbol{g}(x_1).
\end{equation*}

\begin{figure}[!t]
    \centering
    \includegraphics[width=0.485\textwidth]{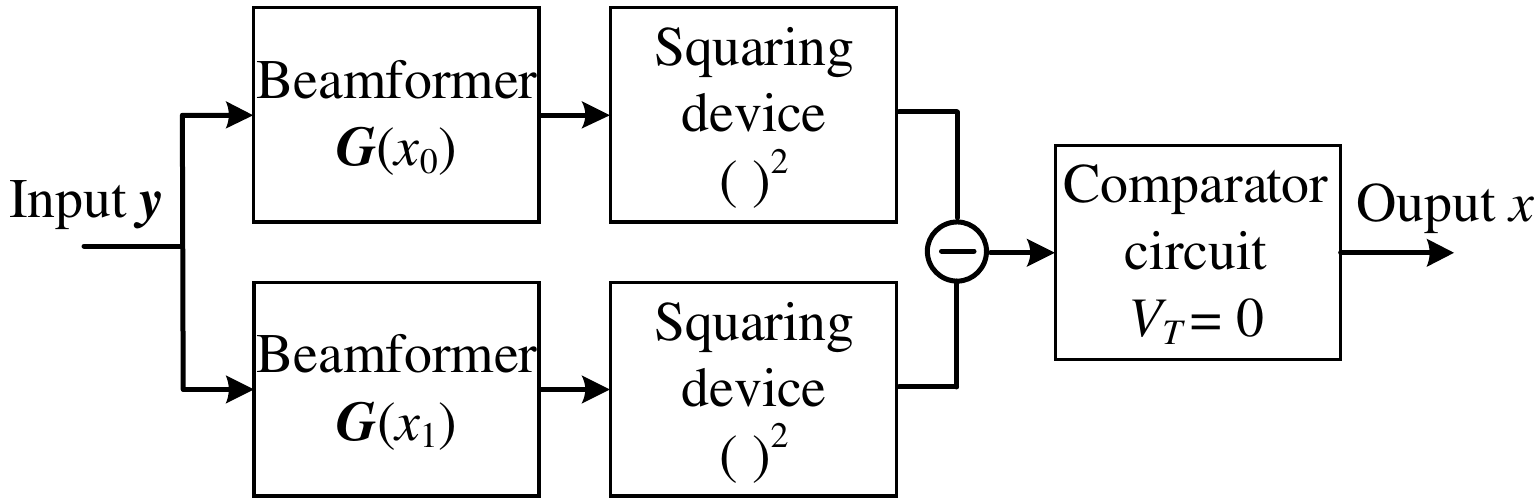}
    \caption{Diagram of the optimum receiver with two beamformers}
     \label{fig:Optimum receiver}
\end{figure}

In Eq.~\eqref{eq:testStatistic}, the test statistic of the optimum multi-antenna receiver of AmBC system $z$ and its decision threshold 0 are given. 
The structure of the optimum receiver is shown in Fig.~\ref{fig:Optimum receiver}. In particular, at one time instant, the received signal is split up into two streams. Each of the streams goes through a beamformer and outputs the beamformed signal written as $\boldsymbol{G}(x_0)^H\boldsymbol{y}$ and $\boldsymbol{G}(x_1)^H\boldsymbol{y}$, respectively. The receiver measures the energy of beamformed signals and compares the energy difference with threshold 0 which does not depend on signals.

The optimum receiver is built upon the fact that when the BD transmits different signal, the direction of the composite channel impinging at the receiver changes. Backscatter signal is detected by measuring the direction of the composite channel at the multi-antennas receiver. Hence, the receiver is effective for the scenario where the directions $\boldsymbol{g}_0$ and $\boldsymbol{g}_1$ are separable. 

In the sequel, bit error rate (BER)-performances of the optimum receiver for constant-power ambient signal and Gaussian-distributed ambient signal are derived.

% \begin{remark}
% It is worthy pointing out that the derived optimum receiver owns a more generalized structure than the likelihood-ratio-based detector (LRD) in Guo \emph{et. al}~\cite{Huayan2019} since we have no prior information about ambient source signal and the BD can perform general binary modulation. In other words, when the RF source signal follows Gaussian distribution and the BD uses OOK modulation, the optimum receiver in Eq.~\eqref{eq:testStatistic} is the same as LRD.
% \end{remark}

\subsection{Detection performance for constant-power $s$}
In this subsection, the performance of the optimum receiver is analyzed for unknown deterministic ambient signal with constant amplitude, for instance, phase-shift-keying-modulated signal and frequency-shift-keying-modulated signal.

In order to investigate the distribution of the test statistic $z$ in Eq.~\eqref{eq:testStatistic},  let us first denote
\begin{equation*}
    \boldsymbol{M} =  \left[\left( \boldsymbol{I} - \frac{ \boldsymbol{g}_1^{} \boldsymbol{g}_1^H }{\|\boldsymbol{g}_1^{}\|^2} \right) - \left(\boldsymbol{I} - \frac{\boldsymbol{g}_0^{} \boldsymbol{g}_0^H }{\|\boldsymbol{g}_0^{}\|^2}\right)  \right] = \frac{\boldsymbol{g}_0^{} \boldsymbol{g}_0^H }{\|\boldsymbol{g}_0^{}\|^2} -\frac{ \boldsymbol{g}_1^{} \boldsymbol{g}_1^H }{\|\boldsymbol{g}_1^{}\|^2} .
\end{equation*}

\begin{proposition}
The matrix $\boldsymbol{M}$ is a rank-2 indefinite matrix with eigenvalues
\begin{equation*}
\kappa_1 =  \kappa = \sqrt{1 - \frac{|\boldsymbol{g}_0^H \boldsymbol{g}_1^{}|^2}{\|\boldsymbol{g}_1^{}\|^2 \|\boldsymbol{g}_1^{}\|^2}} = - \kappa_2,
\end{equation*}
and corresponding eigenvectors \begin{equation}\label{eq:eigenvectorM}
    \boldsymbol{u}_1 = \frac{\tilde{\boldsymbol{u}}_1}{\|\tilde{\boldsymbol{u}}_1\|} ~~~ \boldsymbol{u}_2 = \frac{\tilde{\boldsymbol{u}}_2}{\|\tilde{\boldsymbol{u}}_2\|}
\end{equation}
where
\begin{subequations}
\begin{align*}
&\begin{aligned}
    \tilde{\boldsymbol{u}}_1 =& -\left(\frac{\sqrt{\|\boldsymbol{g}_0^{}\|^2\|\boldsymbol{g}_1^{}\|^2 - |\boldsymbol{g}_0^H\boldsymbol{g}_1^{}|^2} + \|\boldsymbol{g}_0^{}\|\|\boldsymbol{g}_1^{}\|}{\boldsymbol{g}_1^H\boldsymbol{g}_0^{}}\right) \frac{\boldsymbol{g}_0^{}}{\|\boldsymbol{g}_0^{}\|} \\
    &~~~+ \frac{1}{\sqrt{\|\boldsymbol{g}_1^{}\|^2 - \frac {|\boldsymbol{g}_0^H\boldsymbol{g}_1^{}|^2}{\|\boldsymbol{g}_0^{}\|^2} }} \left( \boldsymbol{I} - \frac{\boldsymbol{g}_0^{} \boldsymbol{g}_0^H }{\|\boldsymbol{g}_0^{}\|^2} \right)\boldsymbol{g}_1^{}   ,
    \end{aligned}  \\
&\begin{aligned}
    \tilde{\boldsymbol{u}}_2 =& \left(\frac{\boldsymbol{g}_0^H\boldsymbol{g}_1^{}}{\sqrt{\|\boldsymbol{g}_0^{}\|^2\|\boldsymbol{g}_1^{}\|^2 - |\boldsymbol{g}_0^H\boldsymbol{g}_1^{}|^2} + \|\boldsymbol{g}_0^{}\|\|\boldsymbol{g}_1^{}\|}\right) \frac{\boldsymbol{g}_0^{}}{\|\boldsymbol{g}_0^{}\|} \\
    &~~~+ \frac{1}{\sqrt{\|\boldsymbol{g}_1^{}\|^2 - \frac {|\boldsymbol{g}_0^H\boldsymbol{g}_1^{}|^2}{\|\boldsymbol{g}_0^{}\|^2} }} \left( \boldsymbol{I} - \frac{\boldsymbol{g}_0^{} \boldsymbol{g}_0^H }{\|\boldsymbol{g}_0^{}\|^2} \right)\boldsymbol{g}_1^{} .
    \end{aligned}
    \end{align*}
\end{subequations}
\end{proposition}

\begin{IEEEproof}
See Appendix.\ref{appendix:matrixM}.
\end{IEEEproof}

Taking the eigendecomposition of $\boldsymbol{M}$, the test statistic $z$ can be further written as
 \begin{equation}\label{eq:testStatistic_difference}
 \begin{aligned}
  z = \boldsymbol{y}^H \boldsymbol{M} \boldsymbol{y}=  \kappa \left( \underbrace{ |\boldsymbol{u}_1^H\boldsymbol{y}|^2}_{t} - \underbrace{ |\boldsymbol{u}_2^H\boldsymbol{y})|^2}_{r}\right)  \mathrel{\substack{\mathcal{H}_0\\ \gtrless \\ \mathcal{H}_1}} 0 
 \end{aligned} 
 \end{equation}
where $t$ and $r$ represented by quadratic forms are two independent non-central chi-square distributed variables, both with degrees of freedom 2 and variance $1/2$, but with different non-centrality parameters given by $\theta_t = 2|s|^2|\boldsymbol{u}_1^H \boldsymbol{g}|^2$ and $\theta_r = 2|s|^2|\boldsymbol{u}_2^H \boldsymbol{g}|^2$, respectively~\cite[ch.~2]{Proakis2001}.

The exact density of the difference of two independent noncentral chi-square variables has bee investigated in work \cite{Provost1996}. However, it is difficult to evaluate since the infinite summations of the closed-form PDF do not converge.
Therefore, alternatively, we rewrite Eq.~\eqref{eq:testStatistic_difference} as a ratio 
\begin{equation*}
     \zeta \triangleq  \frac{|\boldsymbol{u}_1^H\boldsymbol{y}|^2}{|\boldsymbol{u}_2^H\boldsymbol{y}|^2} \mathrel{\substack{\mathcal{H}_0\\ \gtrless \\ \mathcal{H}_1}} 1 .
\end{equation*}

\begin{figure}[!t]
	\centering
	\includegraphics[width=0.465\textwidth]{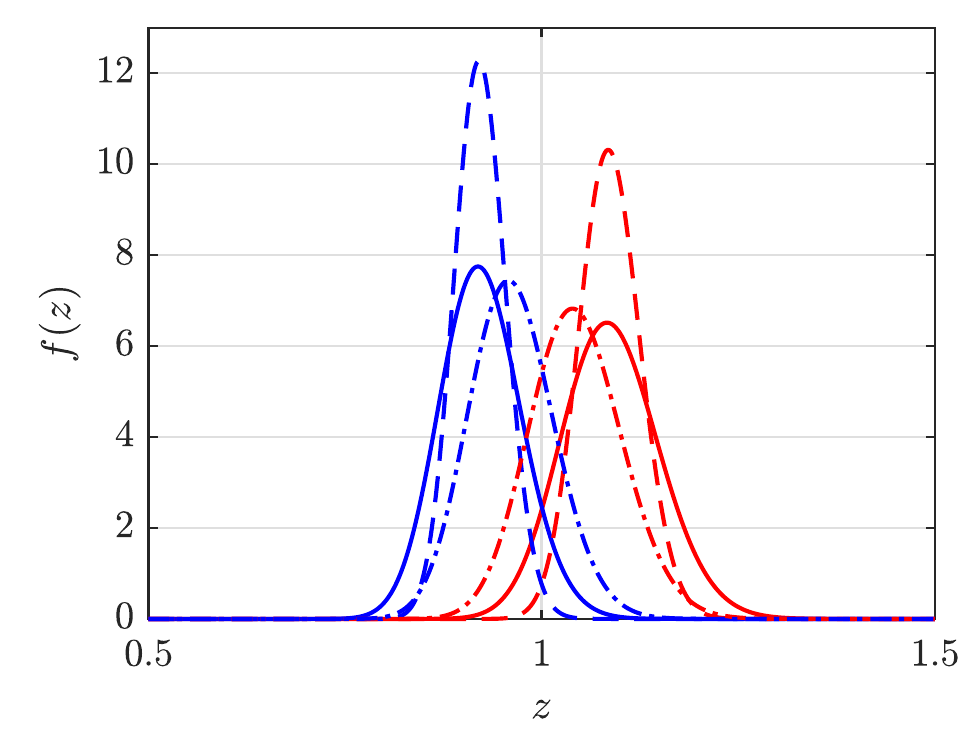}
	\caption{Probability density functions of test statistic $z$ under two hypotheses (\protect\redline~ for $x = x_0$ and \protect\blueline~ for $x = x_1$) for $N_r = 16$, $d_0 = 80\lambda$ and $d_1 = 4\lambda$ with different line styles representing different SNR of legacy system $\gamma$ and BD signal modulation: $\gamma = 22$ and BPSK (solid), $\gamma = 26$ and BPSK (dashed), $\gamma = 22$ and OOK (dash-dotted) }
	\label{fig:ncfpdf}
\end{figure}

By definition, $\zeta$ follows the doubly non-central F distribution. When given a transmitted backscatter signal $x$, the probability density function (PDF) is 
\begin{equation*}
    \begin{aligned}
        f(\zeta|x) = \sum_{i=0}^{\infty} \sum_{j=0}^{\infty} 
        &\frac{\exp\{-\frac{|s|^2}{2}(|\boldsymbol{u}_1^H \boldsymbol{g}|^2 + |\boldsymbol{u}_2^H \boldsymbol{g}(x)|^2)\} }{i!j!} \times \\  &~~ \left(\frac{|s|^2|\boldsymbol{u}_1^H \boldsymbol{g}(x)|^2}{2}\right)^i  \left(\frac{|s|^2|\boldsymbol{u}_2^H  \boldsymbol{g}(x)|^2}{2}\right)^j  \times \\ &~~~~\frac{\zeta^i (\zeta+1)^{-2-i-j}}{B(i+1,j+1)}  ,
    \end{aligned}
\end{equation*}
where $B(a,b)$ denotes the Beta function~\cite[ch.~4]{Johnson1995}.
The cumulative distribution function (CDF) can be expressed as
\begin{equation*}
    \begin{aligned}
        F(\zeta|x) = \sum_{i=0}^{\infty} \sum_{j=0}^{\infty} &\frac{\exp\{-\frac{|s|^2}{2}(|\boldsymbol{u}_1^H \boldsymbol{g}(x)|^2 + |\boldsymbol{u}_2^H \boldsymbol{g}(x)|^2)\} }{i!j!} \times \\ &~~ \left(\frac{|s|^2|\boldsymbol{u}_1^H \boldsymbol{g}(x)|^2}{2}\right)^i  \left(\frac{|s|^2|\boldsymbol{u}_2^H \boldsymbol{g}(x)|^2}{2}\right)^j \times \\&~~~~ \Bar{B}_{\frac{\zeta}{\zeta+1}}(i+1,j+1),
    \end{aligned}
\end{equation*}
where $\Bar{B}_x(a,b)$ denotes the incomplete beta function. Hence, the error probability for the desired BD signal is calculated as
\begin{equation}\label{eq:ORErrorProbability}
    P_e = \frac{1}{2}\left[F(\zeta=1|x=x_0) + (1 - F(\zeta=1|x=x_1))\right] .
\end{equation}
The exact calculation for doubly non-central F distribution is not implemented MATLAB library and runs inefficiently in computational softwares such as Mathematica and R, especially in the case of large non-centrality parameters. There are two methods of approximating the exact PDF. One is the saddle point approximation provided in~\cite[sec.~10.2]{Paolella2007} while the other one is using singly non-central F distribution as the approximation~\cite{Patnaik}. Comparing two approximations, the saddle point approximation performs accurately and its running speed is considerably improved. In Fig.~\ref{fig:ncfpdf}, two examples of doubly non-central F PDFs are illustrated for different SNR of legacy system $\gamma$ and different BD signal modulations.

\subsection{Detection performance for stochastic $s$}
In this subsection, the performance analysis of the optimum receiver is developed to the case of Gaussian-distributed ambient signal $s$. We assume the ambient signal $s \sim \mathcal{CN}(0, \sigma_s^2)$ is a zero-mean complex Gaussian random variable with variance $\sigma_x^2$. The received signal given the backscatter signal $x$ is $\boldsymbol{y} \sim \mathcal{CN}\{\boldsymbol{0}, \boldsymbol{R}_{\boldsymbol{y}|x}\}$ where the covariance matrix is given by $\boldsymbol{R}_{\boldsymbol{y}|x} = \sigma_s^2 \boldsymbol{g}(x)\boldsymbol{g}(x)^H + \boldsymbol{I}$. Hence, the received signal given $x$ can be represented as  $\boldsymbol{y} = \boldsymbol{R}_{\boldsymbol{y}|x}^{1/2} \boldsymbol{v}$ where $\boldsymbol{v} \sim \mathcal{CN}\{\boldsymbol{0},\boldsymbol{I}\}$ is a standard circularly symmetric complex Gaussian random vector. 

In order to obtain the distribution of $z$, we rewrite the test statistic in Eq.~\eqref{eq:testStatistic} as 
\begin{equation*}
\begin{aligned}
        z|x &= \boldsymbol{y}^H \boldsymbol{M} \boldsymbol{y} \\
 &= \boldsymbol{v}^H \boldsymbol{R}_{\boldsymbol{y}|x}^{1/2} \boldsymbol{M} \boldsymbol{R}_{\boldsymbol{y}|x}^{1/2} \boldsymbol{v} .
\end{aligned}
\end{equation*}
We observe that the test statistic is the central quadratic forms of which distribution is depending on the eigenvalues of matrix $\boldsymbol{H}|x = \boldsymbol{R}_{\boldsymbol{y}|x}^{1/2} \boldsymbol{M} \boldsymbol{R}_{\boldsymbol{y}|x}^{1/2}$. Adopting the same approach presented in Appendix of our previous work~\cite{Xiyu2020CoherentReceiver}, we are ready to know that the test statistic follows Asymmetric Laplace distribution~\cite[ch.~3]{Kotz2001} with its PDF and CDF expressed as
\begin{subequations}
\begin{align*}
    F(\zeta| x) &= \begin{cases} 
      -\frac{\lambda_1(x)}{\lambda_2(x) - \lambda_1(x)}  e^{-\frac{\zeta}{\lambda_1(x)}} & \zeta < 0 \\
      1 - \frac{ \lambda_2(x)}{\lambda_2(x) - \lambda_1(x)} e^{-\frac{\zeta}{\lambda_2(x)}} & \zeta \geq 0 
  \end{cases} , \\
    f(\zeta|x) &=  
      \frac{1}{\lambda_2(x) - \lambda_1(x)} 
      \begin{cases} 
      e^{-\frac{\zeta}{\lambda_1(x)}} & \zeta < 0 \\
      e^{-\frac{\zeta}{\lambda_2(x)}} & \zeta \geq 0  
      \end{cases} ,
  \end{align*}
\end{subequations}
respectively, where $\lambda_{\ell}(x), \ell \in\{1, 2\}$ are eigenvalues of matrix $\boldsymbol{H}|x$, written as
\begin{subequations}
    \begin{align*}
    \begin{aligned}
        \lambda_{\ell}(x_0^{}) = &  \frac{\sigma_s^2 \left(\|\boldsymbol{g}_0^{}\|^2 \|\boldsymbol{g}_1^{}\|^2 - |\boldsymbol{g}_0^H \boldsymbol{g}_1^{}|^2 \right)}{2\| \boldsymbol{g}_1^{}\|^2} \\&
         \begin{aligned}
        ~~+ \frac{(-1)^\ell}{2}
        \Bigg[  \Bigg(\frac{\sigma_s^2 \left(\|\boldsymbol{g}_0^{}\|^2 \|\boldsymbol{g}_1^{}\|^2 - |\boldsymbol{g}_0^H \boldsymbol{g}_1^{}|^2 \right)}{\| \boldsymbol{g}_1^{}\|^2} +2 \Bigg)^2   
        - \frac{4|\boldsymbol{g}_0^H\boldsymbol{g}_1^{}|^2}{\|\boldsymbol{g}_0^{}\|^2 \|\boldsymbol{g}_1^{}\|^2} \Bigg]^{\frac{1}{2}} ,
        \end{aligned} \end{aligned}
        \\
        \begin{aligned}
        \lambda_{\ell}(x_1^{}) = &  \frac{\sigma_s^2 \left(|\boldsymbol{g}_0^H \boldsymbol{g}_1^{}|^2 - \|\boldsymbol{g}_0^{}\|^2 \|\boldsymbol{g}_1^{}\|^2  \right)}{2\| \boldsymbol{g}_0^{}\|^2} \\&
         \begin{aligned}
        ~~+ \frac{(-1)^\ell}{2}
        \Bigg[  \Bigg(\frac{\sigma_s^2 \left(\|\boldsymbol{g}_0^{}\|^2 \|\boldsymbol{g}_1^{}\|^2 - |\boldsymbol{g}_0^H \boldsymbol{g}_1^{}|^2 \right)}{\| \boldsymbol{g}_0^2\|^2} +2 \Bigg)^2
        - \frac{4|\boldsymbol{g}_0^H\boldsymbol{g}_1^{}|^2}{\|\boldsymbol{g}_0^{}\|^2 \|\boldsymbol{g}_1^{}\|^2} \Bigg]^{\frac{1}{2}} .
        \end{aligned}   \end{aligned}
    \end{align*}
\end{subequations}
Then, the error probability of the optimum receiver under the case of Gaussian-distributed ambient signal $s$ is given by
\begin{equation}\label{eq:pe_Gaussian_s}
    p_e^{} =  \frac{1}{2} \left [\int_{-\infty}^{0} f(\zeta|x_0^{}) d\zeta  +\int_{0}^{\infty} f(\zeta|x_1^{}) d\zeta \right] .
\end{equation}

The optimum receiver comprising two beamformers can be implemented in both analog domain or digital domain. In the analog domain, the received signal is spilt up into two streams and perform beamforming separately, which cause 3 dB loss of SNR unless using a power amplifier at the receiver. When it comes to the digital domain, the received signal can be copied without power degradation, though an analog-to-digital converter (ADC) is obligatory in circuit. However, power amplifier and ADC cause complex receiver structure. 
In the following, we study the practical implementation of the optimum receiver by proposing a simplified version of it, and providing approximations to the beamformers.

\section{Receiver Implementation}\label{sec:pracImplementation}
In this section, a simplified version of the optimum receiver is proposed and its receiver operating characteristic (ROC) is given. Then, approximations to the required beamformers for implementing the receivers are provided.

\subsection{Simplified receiver and its receiver operating characteristic}\label{sec:oneBReceiver}
Inspired by the energy comparison in Eq.~\eqref{eq:energyComparison}, we propose a simplified receiver by considering only one beamformer. The test statistic of the simplified receiver is expressed as
\begin{equation} \label{eq:signalEnergy}
\begin{aligned}
 z_s^{}  = \boldsymbol{y}^H \boldsymbol{G}(x_0)\boldsymbol{y} = \left\{ \begin{array}{lcc}
    \boldsymbol{n}^H \boldsymbol{G}(x_0) \boldsymbol{n}~,  &  \mathcal{H}_0\\
    \boldsymbol{y}(x_1)^H \boldsymbol{G}(x_0) \boldsymbol{y}(x_1) ~.  &  \mathcal{H}_1
 \end{array}
 \right.
 \end{aligned}
\end{equation}
We observe that $z_s^{}$ measures the energy of beamformed signal $\boldsymbol{G}(x_0)\boldsymbol{y}$, which boils down to an energy detector testing for presence of signal against absence of signal. 

An energy detector makes the decision through comparing the output energy with a predefined threshold which is selected based on the decision criterion of the hypothesis testing.  If the prior statistical information about hypothesis testing is available, the \emph{Bayesian} criterion can be used to calculate posterior probabilities given received signal and determine the decision threshold. However, in practice, AmBC receivers have little prior statistical information about the backscatter signal. In such case, we resort to the \emph{Neyman-Pearson} criterion which constrains the probability of false-alarm and then seeks a suitable decision threshold that maximizes the probability of detection. 

For the simplified receiver, we denote the probability of false-alarm as $P_f \triangleq \mathrm{Pr}\{\boldsymbol{y}\boldsymbol{G}(x_0)\boldsymbol{y}>V_T|\mathcal{H}_0\}$ and the probability of detection as $P_d \triangleq \mathrm{Pr}\{\boldsymbol{y}\boldsymbol{G}(x_0)\boldsymbol{y}>V_T|\mathcal{H}_1\}$ where $V_T$ is the decision threshold. 
The detection problem stated in Eq.~\eqref{eq:signalEnergy} becomes a typical energy detection problem~\cite{Urkowitz1967, Kostylev2000,Digham2007}. The seminal work~\cite{Urkowitz1967} done by Urkowitz analyzes the energy detection problem of a deterministic signal with unknown structure in white Gaussian noise. Kostylev in work~\cite{Kostylev2000} develops the problem to the case of a signal with random amplitude. The closed-form expressions for the probability of detection is investigated in~\cite{Digham2007}. These aforementioned researches calculate the signal energy over a time interval, i.e., explore the time complexity. In our case, we explore the spatial complexity and assume the unknown ambient signal is deterministic. 

Next, we study the performance of this energy detector which is evaluated by the ROC that shows the variation of probability of false-alarm $P_f$ and probability of detection $P_d$. For this purpose, distributions of the test statistic $z_s^{}$ are investigated under two hypotheses. 
Observed from Eq.~\eqref{eq:signalEnergy}, the test statistic $z_s^{}$ is a quadratic form determined by a singular idempotent matrix $\boldsymbol{G}(x_0)$. Distributions of $z_s^{}$ under two hypotheses are defined by the characteristic function of $z_s^{}$. Turin in~\cite{Turin1960} derives a general form of characteristic function of quadratic form for complex multivariate Gaussian case as
\begin{equation*}
    \begin{aligned}
        \psi(\omega) = \frac{\exp\left\{-\Bar{\boldsymbol{y}}^H \left( \boldsymbol{I} - \left(\boldsymbol{I}-j\omega\boldsymbol{\Sigma}\boldsymbol{G}(x_0) \right)^{-1} \right) \Bar{\boldsymbol{y}}\right\}}{\mathrm{det}\left[\boldsymbol{I} - j\omega \boldsymbol{\Sigma}\boldsymbol{G}(x_0)\right]},
    \end{aligned}
\end{equation*}
where $\Bar{\boldsymbol{y}}$ and $\boldsymbol{\Sigma}$ are the mean and covariance matrix of $\boldsymbol{y}$.

When $\mathcal{H}_0$ is true, the energy of beamformed signal is given by
\begin{equation*}
    z_s^{} = \boldsymbol{n}^H \boldsymbol{G}(x_0) \boldsymbol{n}, 
\end{equation*}
which yields its characteristic function, written as
\begin{equation}\label{eq:CFofH0}
    \psi(\omega) = \mathrm{det}|\boldsymbol{I} - j\omega \boldsymbol{G}(x_0)|^{-1} = (1-j\omega)^{1-N_r} .
\end{equation}
%In order to obtain the determinant of matrix $\boldsymbol{I} - j\omega \boldsymbol{G}(x_0)$, we analyze its eigendecomposition . The normalized vector $\boldsymbol{g}_0/\|\boldsymbol{g}_0\|$ is one eigenvector corresponding to eigenvalue 1. And vectors that are perpendicular to $\boldsymbol{g}_0/\|\boldsymbol{g}_0\|$ are eigenvectors associated with eigenvalue $1-j\omega$ whose multiplicity is $N_r-1$. Consequently, we have $\mathrm{det}[\boldsymbol{I} - j\omega \boldsymbol{G}(x_0)] = (1-j\omega)^{N_r-1}$.
Since there is a bijection between characteristic function and probability distribution, the characteristic function Eq.~\eqref{eq:CFofH0} completely defines that $z_s^{}$ is a chi-square-distributed random variable with $2(N_r -1)$ degrees of freedom. The coefficient 2 arises from the fact that each component of $z_s^{}$ is complex value of which common variance equals to  $1/2$ per real and imaginary components.  Then, the PDF of $z_s^{}$ can be written as~\cite[sec.~2.3]{Proakis2001}
\begin{equation*}
    f_{\chi}(z_s^{}) = \frac{1}{\Gamma(N_r-1,0)}z_s^{N_r-2}\exp\{-z_s^{}\}, 
\end{equation*}
where $\Gamma(\cdot,\cdot)$ is the \emph{upper incomplete gamma} function~\cite[sec.~6.5]{Abramowitz}.

Similarly, when $\mathcal{H}_1$ is true, the output is written as
\begin{equation*}
\begin{aligned}
    z_s^{} = \boldsymbol{y}(x_1)^H \boldsymbol{G}(x_0) \boldsymbol{y}(x_1) ,
\end{aligned}
\end{equation*}
where $\boldsymbol{y}(x_1) \sim \mathcal{CN}(s\boldsymbol{g}_1, \boldsymbol{I})$.
In this case, the characteristic function gives
\begin{equation*}
    \begin{aligned}
        \psi(\omega) &= \frac{\exp\left\{-(s\boldsymbol{g}_1)^H \left( \boldsymbol{I} - \left( \boldsymbol{I}-j\omega \boldsymbol{G}(x_0) \right)^{-1} \right) (s\boldsymbol{g}_1) \right\}} {\mathrm{det}\left[\boldsymbol{I} - j\omega  \boldsymbol{G}(x_0)\right]}\\
        &= \frac{1}{(1-j\omega)^{N_r-1}} \exp \left\{ \frac{j\omega |s|^2  }{1-j\omega} \left(\|\boldsymbol{g}_1\|^2 - \frac{|\boldsymbol{g}_1^H \boldsymbol{g}_0^{}|^2}{\|\boldsymbol{g}_0^{}\|^2}\right)\right\} ,
    \end{aligned}
\end{equation*}
which implies $z_s^{}$ follows the non-central chi-square distribution with $2(N_r-1)$ degrees of freedom, common variance $1/2$, and non-centrality parameter 
\begin{equation}\label{eq:non-centralityParameter}
\begin{aligned}
    \theta &= |s|^2 \cdot \left(\|\boldsymbol{g}_1\|^2 - \frac{|\boldsymbol{g}_1^H \boldsymbol{g}_0^{}|^2}{\|\boldsymbol{g}_0^{}\|^2}\right)  \\
    &= \gamma \left(\frac{\lambda}{4\pi d_0}\right)^2 \cdot|x_0-x_1|^2   \cdot \frac{\|\boldsymbol{\alpha}\|^2\|\boldsymbol{\beta}\|^2 - |\boldsymbol{\alpha}^H \boldsymbol{\beta}|^2}{\|\boldsymbol{\alpha} + x_0\boldsymbol{\beta}\|^2} .
\end{aligned}
\end{equation}
Then, the PDF of $z_s^{}$ is given by
\begin{equation*}
    f_{\chi'}(z_s^{}) = \left(\frac{z_s^{}}{\theta}\right)^{\frac{N_r-2}{2}}\exp\left\{-(z_s^{}+\theta)\right\}I_{N_r-2}\left(2\sqrt{\theta z_s^{}}\right) ,
\end{equation*}
where $I_v(\cdot)$ is the \emph{$v$-th order modified Bessel function of the first kind}~\cite[sec.~9.6]{Abramowitz}.

With density functions of the test statistic under two hypotheses, the probabilities of phase-alarm and detection for a given threshold $V_T$ are 
\begin{subequations}
    \begin{align*}
    &\begin{aligned}
        P_f &= \mathrm{Pr}\{z_s^{}>V_T|\mathcal{H}_0\} \\
        &= \Bar{\Gamma}(N_r-1,V_T) = \frac{\Gamma(N_r-1,V_T)}{\Gamma(N_r-1,0)} ,
    \end{aligned} \\
    &\begin{aligned}
       P_d &= \mathrm{Pr}\{z_s^{}>V_T|\mathcal{H}_1\} \\
       &= Q_{N_r-1}(\sqrt{2\theta},\sqrt{2V_T}),
       \end{aligned}
    \end{align*}
\end{subequations}
where $\Bar{\Gamma}(\cdot,\cdot)$ is the \emph{regularized upper incomplete gamma function}~\cite[sec.~6.5]{Abramowitz},
and $Q_{N_r-1}(\cdot, \cdot)$ is the \emph{generalized Marcum Q-function}. The decision threshold can be decided from the $P_f$ using the inverse function of $\Bar{\Gamma}(\cdot,\cdot)$, expressed as
\begin{equation*}
    V_T =  \Bar{\Gamma}^{-1} (N_r-1,P_f).
\end{equation*}
Hence, the probability of detection can be represented in terms of $P_f$
\begin{equation}\label{eq:Pd}
    P_d = Q_{N_r-1}\left(\sqrt{2\theta}, \sqrt{2 \Bar{\Gamma}^{-1} (N_r-1,P_f)}\right),
\end{equation}
which is determined by the non-centrality parameter $\theta$. 
Then, the error probability of equal-probability BD signal can be written as
\begin{equation}\label{eq:SORErrorProbability}
    P_e = \frac{1}{2} \left[ P_f + (1 - P_d)\right]. 
\end{equation}

\begin{figure}[!t]
    \centering
    \includegraphics[width=0.485\textwidth]{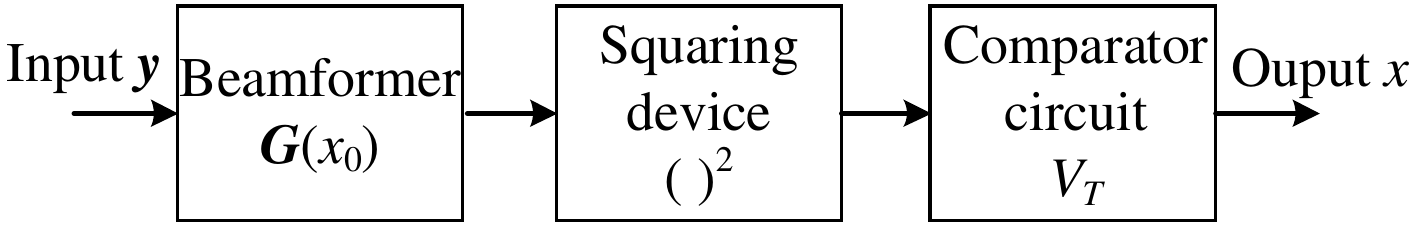}
    \caption{Diagram of the simplified optimum receiver}
    \label{fig:receiverStructure}
\end{figure}

Finally, the structure of the simplified receiver is shown in Fig.~\ref{fig:receiverStructure}. The received signal $\boldsymbol{y}$ going through the beamformer $\boldsymbol{G}(x_0)$ is fed into a squaring device. Then, the obtained signal energy is compared with a selected threshold $V_T$ so as to identify the BD signal to be the null hypothesis or its alternative.  It is noteworthy that the receiver can be implemented in the analog domain, which relaxes the requirement of a high dynamic range and saves extra power for the analog-to-digital converter (ADC) component.

\subsection{Approximations to the beamformers}

The beamformers $\boldsymbol{G}(x_0)$ and $\boldsymbol{G}(x_1)$ required for constructing two receivers are built upon the knowledge of instantaneous channels, i.e., $\boldsymbol{\alpha}$ and $\boldsymbol{\beta}$. 
However, in practice, it is challenging to track down channel conditions of AmBC systems since the legacy system provides little cooperation to AmBC system such that their preamble information is not available at AmBC receivers. On the other hand, a long preamble sequence is required due to the weak backscatter channel even when preamble information is given, which induces the severe energy cost for the BD. In this subsection, we study practical approximations to two beamformers while avoiding estimating the channels $\boldsymbol{\alpha}$ and $\boldsymbol{\beta}$ separately.

Preambles are prepended to the information bits of BD signal in order to obtain the beamformers within one channel coherence time. Two length-$L$ preambles represented by $x[1] =\cdots = x[L] = x_0$ and $x[L+1] = \cdots= x[2L] = x_1$ are used for estimating $\boldsymbol{G}(x_0)$ and $\boldsymbol{G}(x_1)$, respectively. For the simplified optimum receiver, the second preamble can be omitted since only one beamformer is required. Such property conserves BD transmission energy which corresponds to the increase of data rate. In the sequel, we provide three practical methods of obtaining $\boldsymbol{G}(x_0)$ while $\boldsymbol{G}(x_1)$ can be estimated similarly using another preamble.  Let us denote $\boldsymbol{Y}_p = [\boldsymbol{y}_1,\cdots, \boldsymbol{y}_L]$ as the sample matrix of which $l$-th column represents the $l$-th sample during preamble sequence.

\subsubsection{Inverse of sample covariance matrix}
At the AmBC receiver, the sample covariance matrix of received signal is given by
\begin{equation}
\boldsymbol{R}_p = \frac{1}{L} \sum_{l=1}^{L}\{\boldsymbol{y}_l\boldsymbol{y}_l^H\} \approx |s|^2\boldsymbol{g}_0^{} \boldsymbol{g}_0^H + \boldsymbol{I} .
\end{equation}
where the sample covariance matrix asymptotic to the third term as $L$ goes to infinity. 
Then, the inverse of matrix $\boldsymbol{R}_p$ is given by
\begin{equation*}
\begin{aligned}
\boldsymbol{R}_p^{-1} &= \boldsymbol{I} - \frac{|s|^2\|\boldsymbol{g}_0^{}\|^2}{1 + |s|^2 \|\boldsymbol{g}_0\|^2} \frac{\boldsymbol{g}_0^{} \boldsymbol{g}_0^H }{\|\boldsymbol{g}_0^{}\|^2}
%&= \boldsymbol{I} -   \frac{\gamma(4\pi d_0/\lambda)^2}{1 + \gamma(4\pi d_0/\lambda)^2 \|\boldsymbol{g}_0\|^2} \boldsymbol{g}_0^{} \boldsymbol{g}_0^H
\end{aligned}
\end{equation*}
where we use the Sherman–Morrison formula~\cite{Tylavsky1986}. As can be seen, $\boldsymbol{R}_p^{-1} \approx \boldsymbol{G}(x_0)$ holds when $|s|^2\|\boldsymbol{g}_0^{}\|^2$ is large. In other words, it requires large SNR seen by the AmBC receiver. Therefore, this method is constrained by factors including length of preambles and SNR value.

\subsubsection{Singular value decomposition of Rx samples}
It has been investigated in work~\cite{Zhao2018BlindCE} that  $\boldsymbol{g}_0$ is the eigenvector with respect to the largest eigenvalue of the sample covariance matrix. 
However, calculating the covariance matrix increases the computational complexity of the receiver. Therefore, alternatively, we obtain the approximation to $\boldsymbol{g}_0$ by taking singular value decomposition (SVD) of sample matrix $\boldsymbol{Y}_p$ and selecting the left-singular vector with respect to the largest singular value as the estimate.

Specifically, the SVD of sample matrix is $\boldsymbol{Y}_p = \boldsymbol{U}^{}\boldsymbol{\Sigma}^{}\boldsymbol{V}^H$, where the  columns  of  matrices $\boldsymbol{U}$ and $\boldsymbol{V}$ are the left-singular vectors and right-singular vectors of  signal subspace,  respectively,  and $\boldsymbol{\Sigma}$ contains  their corresponding singular values of $\boldsymbol{Y}_p$. Then we take the singular vector with respect to the largest singular value as the estimate of normalized $\boldsymbol{g}_0$, denoting as 
$\hat{\boldsymbol{g}}_0 = \boldsymbol{u}_1 $.
Then this estimate gives our target beamformer $\boldsymbol{G}(x_0) = \boldsymbol{I} - \hat{\boldsymbol{g}}_0^{}\hat{\boldsymbol{g}}_0^H$.

\subsubsection{Power iteration}
Since we are only interested in the eigenvector associated with the dominant eigenvalue, an alternative way of approximating this eigenvector is the power iteration algorithm. It is represented by a recurrence relation
\begin{equation}
    \boldsymbol{v}_{k+1} = \frac{\boldsymbol{A} \boldsymbol{v}_{k}}{\|\boldsymbol{A} \boldsymbol{v}_{k}\|} ,
\end{equation}
where $\boldsymbol{A} = \boldsymbol{Y}_p^{} \boldsymbol{Y}_p^H$ in our case.
The initialization $\boldsymbol{v}_0$ should have a nonzero component in the direction along with the eigenvector. Then, the vector multiplies matrix $\boldsymbol{A}$ and normalized until convergence criterion satisfied. Afterwards, we obtain the approximation of the dominant eigenvector by $\boldsymbol{v}_{k+1}$. 

\subsection{Discussion}
The non-centrality parameter $\theta$ given in Eq.~\eqref{eq:non-centralityParameter} plays a central role in determining the performance of the simplified receiver. It indicates the effective SNR of the backscatter signal after beamforming. The first term is the transmit power of ambient signal. It associates with SNR of legacy system $\gamma$ when Tx-Rx distance $d_0$ is fixed. Second,  the term $|x_0 - x_1|^2$ denotes the difference between two BD signal alphabets, which suggests that adopting BPSK achieves the same detection performance with 6 dB less SNR than adopting OOK~\cite[ch.~4]{Proakis2001}. The third term indicates the angular distance between $\boldsymbol{\alpha}$ and $\boldsymbol{\beta}$. It reaches the maximum value when $\boldsymbol{\alpha}$ and $\boldsymbol{\beta}$ are linearly independent, while goes to 0 given $\boldsymbol{\alpha}$ and $\boldsymbol{\beta}$ are parallel. 

Then, we investigate the performance comparison between the optimum receiver and the simplified receiver. Considering the test statistics of two receivers, the optimum receiver contains two beamformers $\boldsymbol{G}(x_0)$ and $\boldsymbol{G}(x_1)$ whereas the simplified receiver contains one beamformer $\boldsymbol{G}(x_0)$.
It can be observed that the null spaces of two beamformer matrices $\boldsymbol{G}(x_0)$ and $\boldsymbol{G}(x_1)$ are spanned by $\boldsymbol{g}_0$ and $\boldsymbol{g}_1$, respectively. 
The optimum receiver divides the signal space into two subspaces spanned by $\boldsymbol{g}_0$ and $\boldsymbol{g}_1$ and demodulates the backscatter signal through checking the signal space of the received signal. On the contrary, the simplified receiver divides the signal space into one subspace spanned by $\boldsymbol{g}_0$ and its null space. The received signal under $\mathcal{H}_1$ cannot fully fall into the null space of $\boldsymbol{g}_0$. Such mathematical fact leads to the result that the simplified receiver cannot reach the performance of the optimum receiver unless $\boldsymbol{g}_1$ is perpendicular to $\boldsymbol{g}_0$. However, in practical AmBC system, $\boldsymbol{g}_1$ cannot be orthogonal to $\boldsymbol{g}_0$ since the backscatter channel is significantly weaker compared with the direct channel.

With respect to the hardware implementation of the simplified receiver, since it only has one beamformer, only one training sequence is needed at the BD. It consumes less energy which in turn improves the data rate of the backscatter signal. Furthermore, 
several circuit components can be saved for estimating beamformers and for carrying out beamforming at the receiver.  
Therefore, the simplified multi-antenna receiver realizes a low-cost and simple reception of general binary-modulated backscatter signal.

\section{Simulation Results}\label{sec:simulation}
In this section, 
numerical results are provided to evaluate the performance of the proposed receivers. For this purpose,  we first present the numerical evaluations for, respectively, the simplified receiver and the optimum receiver about different parameters to study their impacts. Then, we investigate the performance of two receivers under the circumstances that the BD and the ambient source adopt different modulations. All the results are obtained by averaging over $10^6$ Monte-Carlo realizations. The distances are normalized with respect to wavelength so that the results are carrier-frequency-independent.

We consider the Rx has a linear array with half-wavelength $\lambda/2$ antenna separation. The distance between the Tx antenna and Rx reference antenna is $d_0 = 80\lambda$. Without further notice, the BD is placed at $[(40 - 4/\sqrt{2})\lambda, (4/\sqrt{2})\lambda]$ in the coordinate system shown in Fig.~\ref{fig:systemModel}, i.e., $d_1 = 4\lambda$, which implies the backscatter path endures a 33.7 dB power loss compared with the direct path. Hence, we can approximate the effective SINR of the backscatter signal given the SNR of the legacy system $\gamma$.

\begin{figure}[!t]
	\centering
	\includegraphics[width=0.465\textwidth]{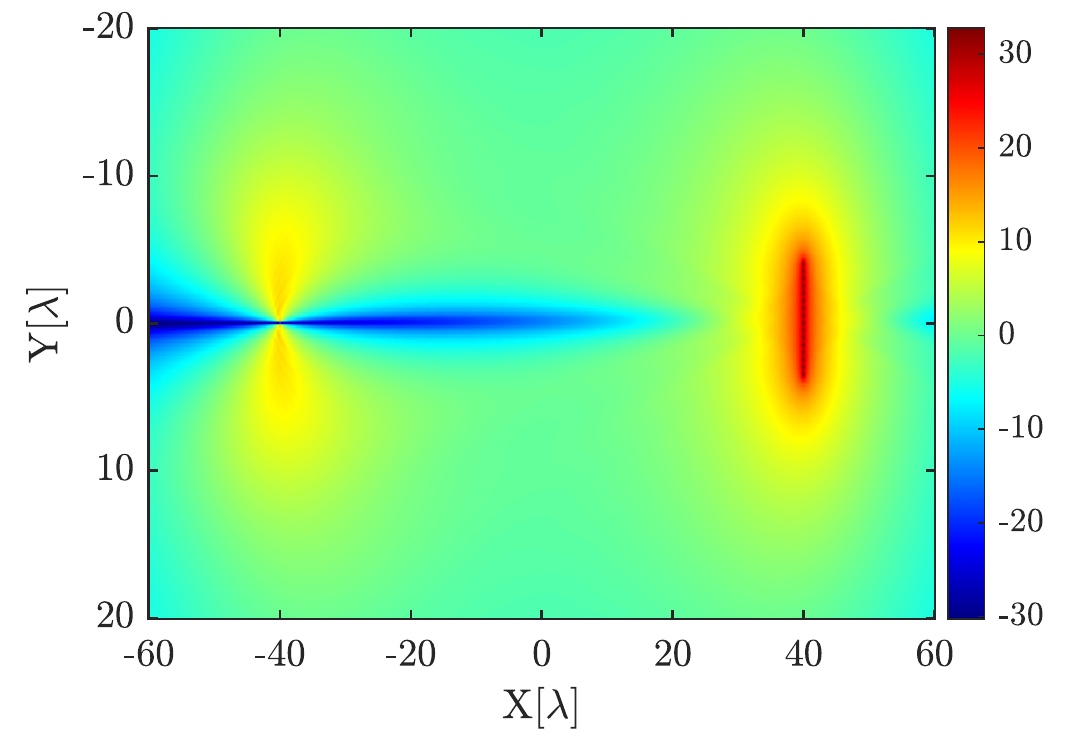}
	\caption{Variation of non-centrality parameter $\theta$ in dB as a function of BD location for $d_0 = 80\lambda$ and legacy system SNR of 28 dB}
	\label{fig:noncentralParam}
\end{figure}

\subsection{Evaluation for the simplified receiver}

As we discussed in Section~\ref{sec:pracImplementation}, the performance of the simplified receiver is decided by the non-centrality parameter $\theta$ which implies the effective SNR of backscatter signal. In Fig.~\ref{fig:noncentralParam}, variation of $\theta$ in dB as a function of BD location for BPSK-modulated backscatter signal, number of antennas $N_r = 16$ and legacy system SNR $\gamma = 28$ dB is shown.  The BD is placed in a $(120 \lambda \times 40 \lambda)$ area in which the Tx and the reference antenna of Rx are placed at $[-40\lambda,0]$ and $[40\lambda,0]$, respectively. Observing from the figure, the non-centrality parameter $\theta$ increases as the BD moves close to either the Tx or the Rx. When BD is located farther away from the Tx and the Rx, the effective SNR of backscatter signal reduced to around 0 dB, which causes poor detection performance. 
Moreover, there exits a null beam along the line between the Tx and the Rx reference antenna which represents a even worse detection performance. 
The reason is that the directions $\boldsymbol{g}_0$ and $\boldsymbol{g}_1$ are inseparable in this area. The result suggests that the BD should not in the null beam but in the close vicinity of Tx and Rx.

\begin{figure*}
	\centering
	\setlength{\tabcolsep}{0pt}
	\begin{tabular}{ccc}
		\subfloat[]{\includegraphics[width=0.32\textwidth]{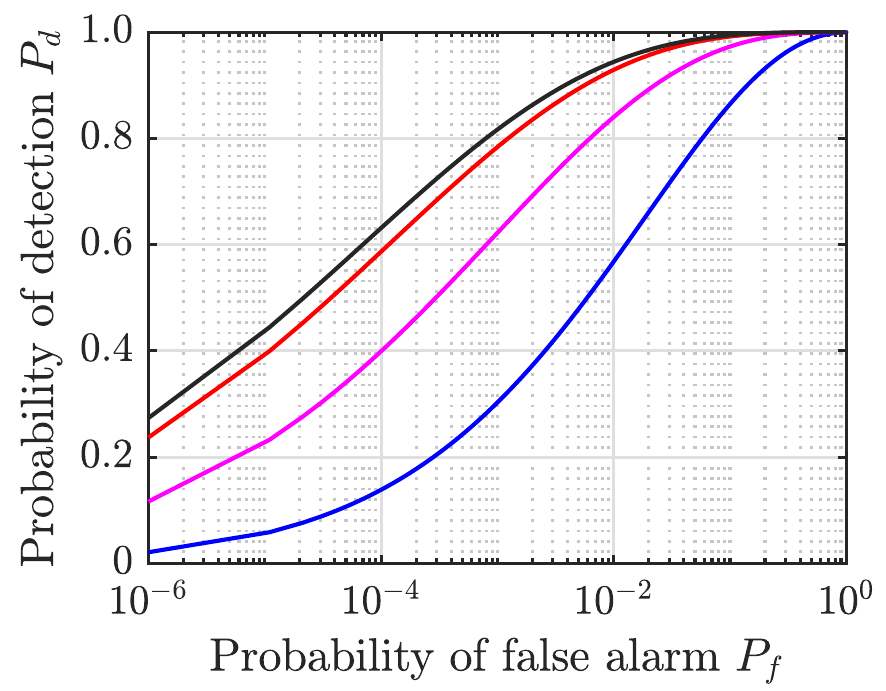}
			\label{fig:Pf_Pd}} &
		\subfloat[]{\includegraphics[width=0.32\textwidth]{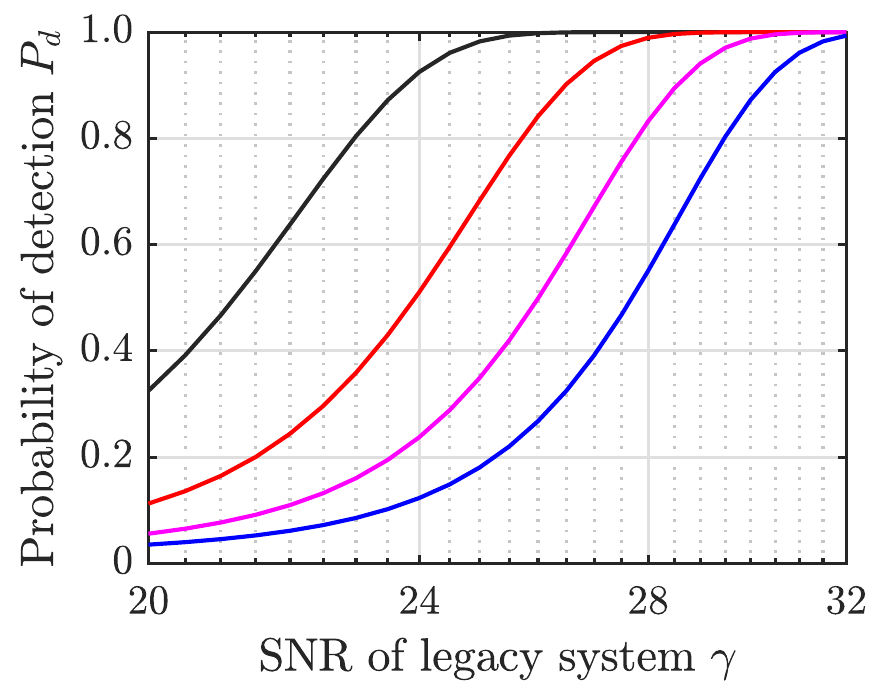} 
			\label{fig:Pf_SNR}} &
		\subfloat[]{\includegraphics[width=0.32\textwidth]{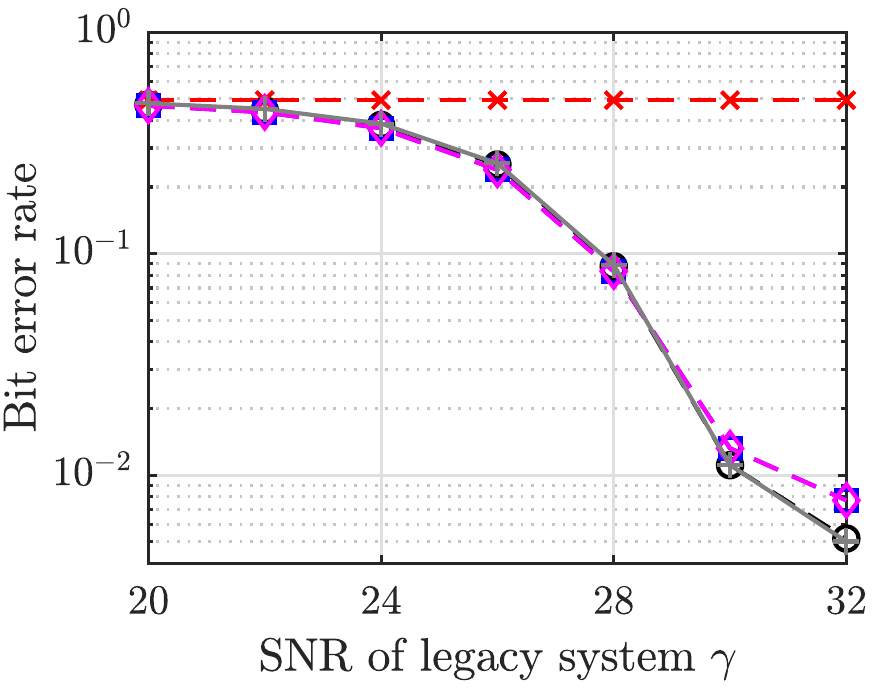}\label{fig:Simplified_BER_SNR}}
	\end{tabular}
	\caption{Numerical evaluation of the simplified receiver. In (a), variation of ROC curves for $\gamma = 28$ dB with different colors representing number of antenna $N_r$: 8 (\protect\blueline), 16 (\protect\magentaline), 24 (\protect\redline), 32 (\protect\blackline).  In (b), variation of $P_d$ as a function of $\gamma$ for $P_f = 10^{-2}$ and $N_r = 16$ with different colors representing distance between the BD and the Rx reference antenna $d_1$: $5\lambda$ (\protect\blueline), $4\lambda$ (\protect\magentaline), $3\lambda$ (\protect\redline), $2\lambda$ (\protect\blackline). In (c), variation of AmBC BER as a function of $\gamma$ for $N_r = 16$, $P_f = 10^{-2}$ with different marker representing: theoretical probability in Eq.~\eqref{eq:SORErrorProbability} (\protect\grayplus), perfect channel conditions (\protect\blackcircle), power iteration (\protect\magentadiamond), SVD (\protect\bluesquare), and inverse of sample covariance matrix (\protect\redcross). }
	\label{fig:SR_evaluation}
\end{figure*}

Performance evaluations for the simplified receiver are shown in Fig.~\ref{fig:SR_evaluation}. In Fig.~\ref{fig:Pf_Pd}, the ROC curves of the receiver for SNR of legacy system $\gamma = 28$ dB and different number of antennas $N_r$ are shown. As $N_r$ increases, the non-centrality parameter in Eq.~\eqref{eq:non-centralityParameter} increases, and thus the probability of detection increases for $P_f \in [1\cdot10^{-6}, 1\cdot10^{0}]$ as implied by Eq.~\eqref{eq:Pd}. When $\gamma$ is 28 dB, $P_f = 10^{-2}$ yields high probability of detection for $N_r$ is 16 or more.  For this specific $P_f$ and $N_r=16$, variation of $P_d$ as a function of $\gamma$ for different $d_1$ values are shown in Fig.~\ref{fig:Pf_SNR}. For a fixed $\gamma$ and $N_r$, the detection probability increases with decreasing distance between the BD and Rx reference antenna as a shorter distance improves the effective SNR of the backscatter signal which in turn boosts the non-centrality parameter. 

In Fig.~\ref{fig:Simplified_BER_SNR}, bit-error-rate (BER)-performances of the simplified receiver as a function of $\gamma$ for $N_r=16$, $P_f = 10^{-2}$ and $d_1 = 4\lambda$ are illustrated. BER-performances of using three estimation methods discussed in Section~\ref{sec:pracImplementation} are compared with the theoretical error probability and BER-performance of perfect channel conditions. As shown, in the high SNR region, the simplified receiver suffers from the error floor issue. For estimation methods, training sequence of length $L = 30$ is utilized. Power iteration and SVD of sample matrix provide decent approximations of the beamformers and have very close performance compared to the ideal case. However, the method of inverse of sample covariance does not work since $\boldsymbol{R}_p^{-1}$ is not approximate to $\boldsymbol{G}(x_0)$ in such SNR region. Therefore, either power iteration and SVD of sample matrix can be used for practical implementation of the receiver.

\subsection{Evaluation for the optimum receiver}
\begin{figure*}
	\centering
	\setlength{\tabcolsep}{0pt}
	\begin{tabular}{ccc}
		\subfloat[]{\includegraphics[width=0.32\textwidth]{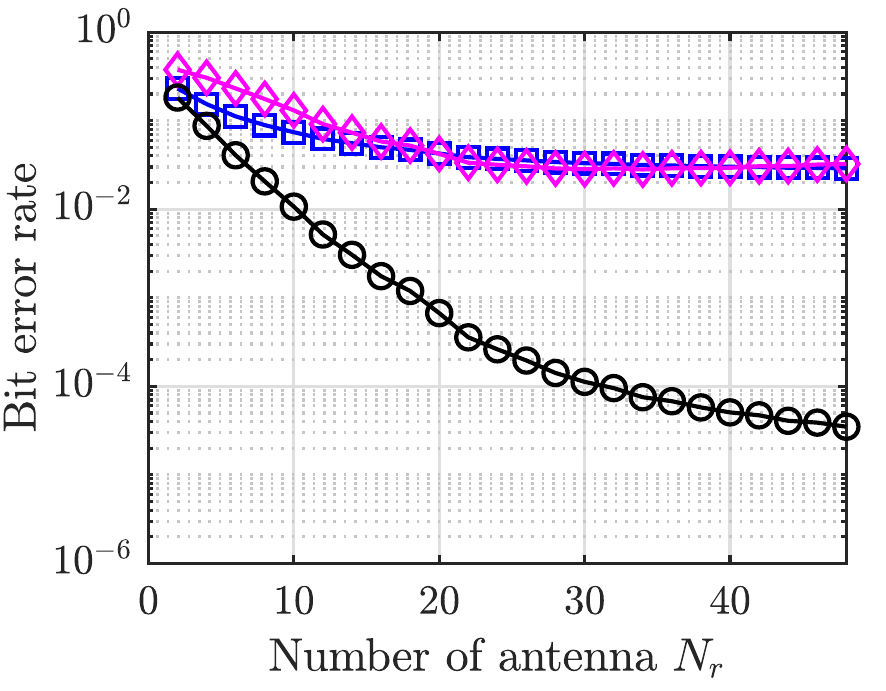}
			\label{fig:BER_lengthPilot}} &
		\subfloat[]{\includegraphics[width=0.32\textwidth]{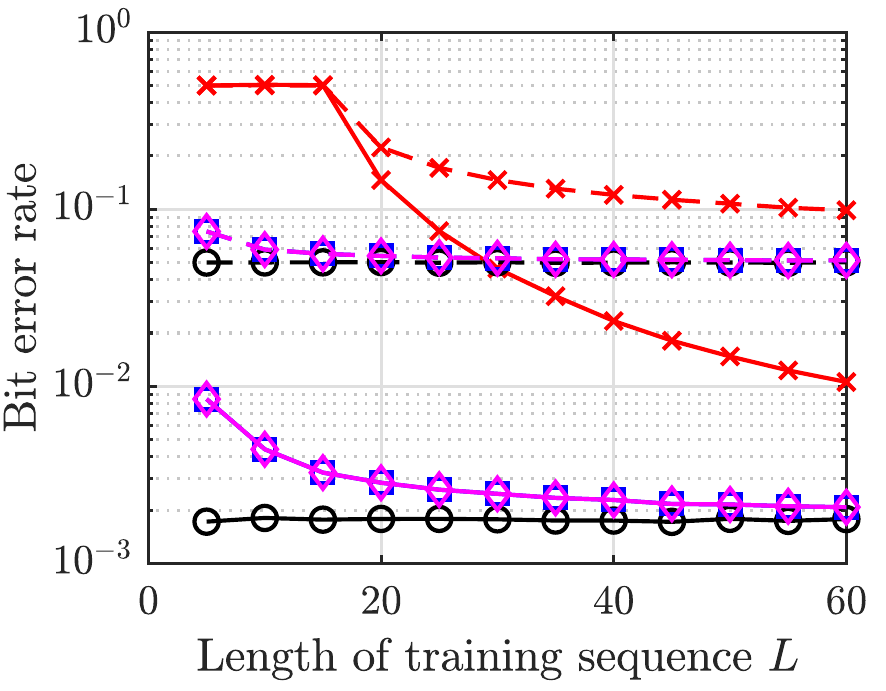}\label{fig:BER_Nr}} &
		\subfloat[]{\includegraphics[width=0.32\textwidth]{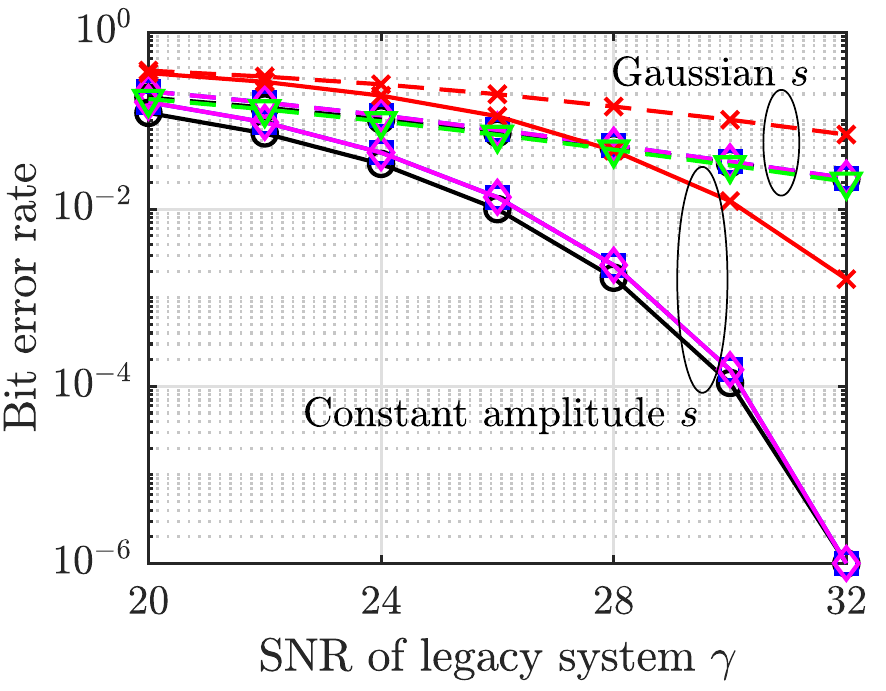}
			\label{fig:BER_SNR}}
	\end{tabular}
	\caption{Numerical evaluation of the optimum receiver. In (a), variation of AmBC BER as a function of $N_r$ for $\gamma = 28$ dB with different markers representing: optimum receiver for ambient signal with constant $|s|^2$ (\protect\blackcircle), optimum receiver with Gaussian-distributed $s$ (\protect\bluesquare), and simplified receiver with constant $|s|^2$ (\protect\magentadiamond).   In (b), variation of AmBC BER as a function of $L$ for $N_r=16$ and $\gamma = 28$ dB.  In (c), variation of AmBC BER as a function of $\gamma$ for $N_r=16$ and $L = 30$. In both (b) and (c), solid lines illustrate the case of constant $|s|^2$ and dashed lines illustrate the case of Gaussian-distributed $s$. Black lines with marker (\protect\blackcircle) represent theoretical error probabilities in Eq.~\eqref{eq:ORErrorProbability} and Eq.~\eqref{eq:pe_Gaussian_s} respectively. Different markers representing different estimation methods: power iteration (\protect\magentadiamond), singular value decomposition (\protect\bluesquare), and inverse of sample variance matrix (\protect\redcross). The line with marker (\protect\greentriangle) in (c) represent the coherent receiver in \cite{Xiyu2020CoherentReceiver}.}
	\label{fig:OR_evaluation}
\end{figure*}
In Fig.~\ref{fig:OR_evaluation}, BER-performances of the optimum receiver with different parameters are shown. Error probabilities of the optimum receiver as a function of $N_r$ with constant-power ambient source signal and Gaussian-distributed ambient signal and of the simplified receiver are compared in Fig.~\ref{fig:BER_Nr}. As can be seen, the improvement of BER-performance is diminishing as $N_r$ becomes larger which is aligned with the performance of the spatial diversity. Considering this result and the size of antenna array, we select the parameter $N_r = 16$ for the simulation.

In Fig.~\ref{fig:BER_lengthPilot}, the effect of varying lengths of training sequence $L$ on the BER-performance of the optimum receiver is shown.  As $L$ increases, error probabilities for all estimation methods approach to the theoretical results. Methods of power iteration and SVD of sample matrix have close performance compared with theoretical error probabilities even given a small value of $L$, and their performances converge when $L$ larger than 30. For the method of inverse of sample covariance matrix $\boldsymbol{R}_p^{-1}$, there still exist performance gap as $L$ becomes larger. It starts to work when $L \geq 16$ as the receiver requires enough samples to calculate the sample covariance matrix for $N_r = 16$. Since a shorter training sequence saves energy of the BD and reduces the computational complexity of the receiver, it is reasonable to choose $L = 30$ for $N_r = 16$.

BER-performances of the optimum receiver as a function of $\gamma$ for constant-power $s$ and Gaussian-distributed $s$ with the choosing parameters are depicted in Fig.~\ref{fig:BER_SNR}. The result clearly shows that the optimum receiver is free of error-floor issue. AmBC system with a constant $|s|^2$ ambient source signal yields better performance than that with Gaussian-distributed $s$.  Furthermore, for the case of Gaussian-distributed $s$, we also provide the coherent receiver studied in our previous work \cite{Xiyu2020CoherentReceiver} as a benchmark. As shown, the proposed optimum receiver performs the same as the coherent receiver. Although the proposed receiver avoids the coherent reception of the backscatter signal, it does not lose the 3-dB gain provided by the coherent scheme \cite[sec.~3.1]{Tse2005}.

\begin{figure}[!t]
	\centering
	\includegraphics[width=0.465\textwidth]{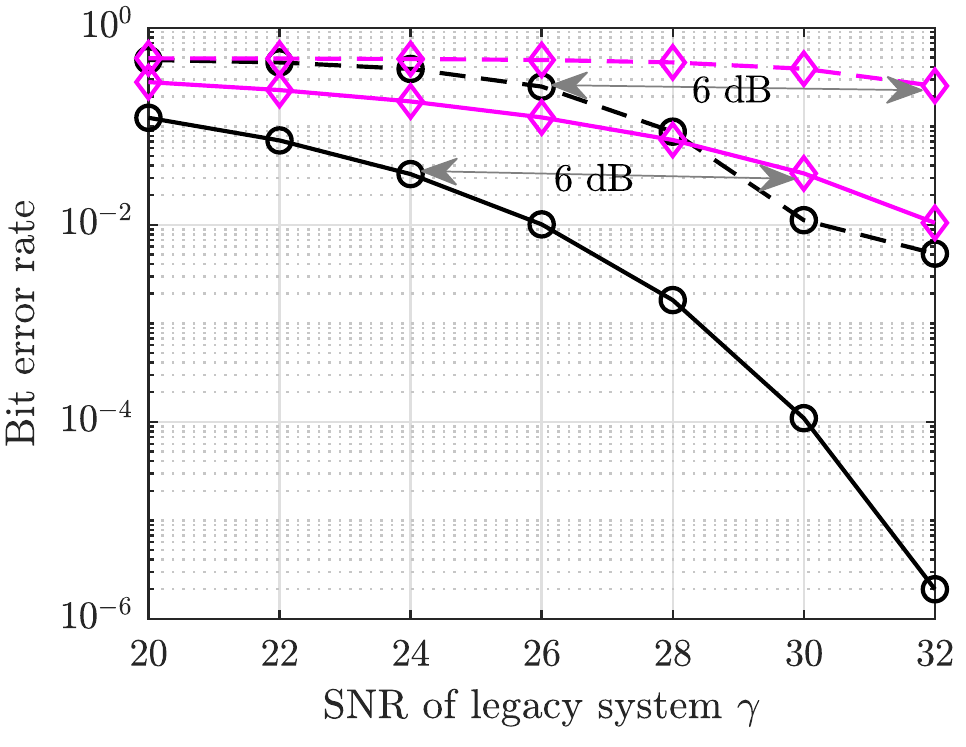}
	\caption{Variation of AmBC BER of the optimum receiver (solid lines) and the simplified receiver (dashed lines) as a function of $\gamma$ for $N_r = 16$ with different markers representing the case of the BD performing BPSK modulation (\protect\blackcircle) and OOK modulation (\protect\magentadiamond).}
	\label{fig:BER_SNR_OOKvsBPSK}
\end{figure}

\subsection{Evaluation for different modulations}
In Fig.~\ref{fig:BER_SNR_OOKvsBPSK}, BER-performances of the optimum receiver and the simplified receiver as a function of $\gamma$ are compared for the BD adopting OOK and BPSK modulation, $N_r = 16$. It can be observed that for a certain receiver, OOK modulation achieves the same BER-performance as BPSK modulation with 6-dB more SNR, which is explained by the term $|x_0 - x_1|^2$ in non-centrality parameter $\theta$ in  Eq.~\eqref{eq:non-centralityParameter}. This result shows that adopting BPSK modulation at the BD helps to improve the detection performance of the backscatter signal. Furthermore, the optimum receiver outperforms the simplified receiver for each modulation with at least 4-dB SNR gain, of which reason has been discussed in subsection \ref{sec:oneBReceiver}.

\begin{figure}[!t]
	\centering
	\includegraphics[width=0.465\textwidth]{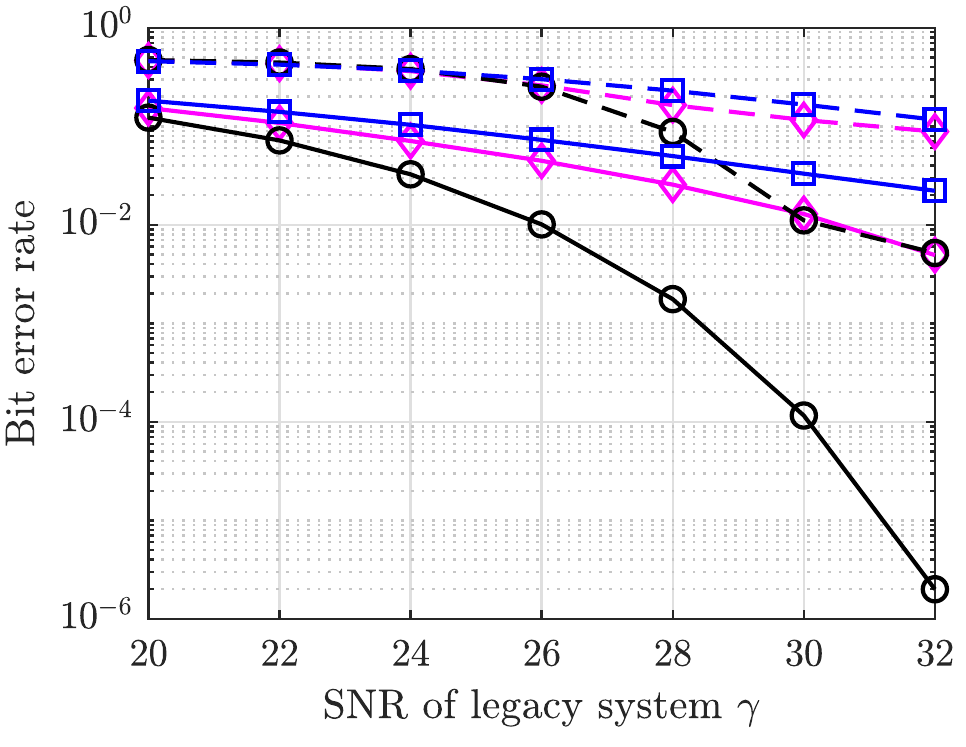}
	\caption{Variation of AmBC BER of the optimum receiver (solid lines) and the simplified receiver (dashed lines) as a function of $\gamma$ for $N_r = 16$ with different markers representing types of ambient signal $s$: constant $|s|^2$ (\protect\blackcircle), 16-QAM $s$ (\protect\magentadiamond), and Gaussian-distributed $s$ (\protect\bluesquare).}
	\label{fig:BER_SNR_CnQamQpsk}
\end{figure}

BER-performances of the proposed receivers as a function of $\gamma$ are compared in Fig.~\ref{fig:BER_SNR_CnQamQpsk} for different ambient source signal, i.e., QPSK-modulated, 16-quadrature amplitude modulation (16-QAM) and Gaussian-distributed source signal. The result in this figure shows that performance of the optimum receiver reaches the highest when $s$ has constant amplitude, i.e., $s \sim$ QPSK. When $s$ has varying amplitudes, 16-QAM-modulated source signal yields slightly better performance compared with Gaussian source signal. Therefore, the proposed receivers can be used with source signal which has deterministic $|s|^2$, for instance, $M$-PSK-modulated source signal, to increases the data rate or the transmission range of the AmBC system. 

\section{Conclusion}\label{sec:conclusion}
In this paper, the optimum multi-antenna receiver for general binary-modulated backscatter signal in AmBC system is studied. The optimum receiver is derived from maximum-a-posteriori criterion with no prior statistical information about ambient signal. 
BER performances of the optimum receiver are derived under Gaussian-distributed ambient signal and unknown deterministic ambient signal with constant amplitude. For facilitating implementation, a simplified receiver with one beamformer is proposed, and efficient estimation methods of required beamformers are provided.  
The work in this paper suggests that the optimum multi-antenna receiver for general binary-modulated backscatter signal leverages the fact that the direction of the received signal is changed by the backscatter signal. The derived receiver owns a simple structure containing two beamformers and the decision threshold is simply 0.  The receiver takes a form of non-coherent receiver but achieves the SNR gain of the coherent reception of backscatter signal. The work in this paper promotes the implementation of a simple and low-cost multi-antenna optimum receiver with high flexibility in AmBC systems. 

\appendices
\section{Proof of Proposition I}
\label{appendix:loglikelihood}
In order to form the log-likelihood testing criterion in Eq.~\eqref{eq:LogMAPcriterionApprox}, one has to obtain the estimate of ambient source signal.  
The asymptotic efficient estimate for the optimum receiver is the $\hat{s}$ that maximizes the likelihood functions, i.e.,
\begin{equation*} 
    \hat{s} = \arg \max_{s} f(\boldsymbol{y}|s) = \arg \max_{s} \sum_{x\in \mathcal{X}} p(x)f(\boldsymbol{y}|x,s).
\end{equation*}
where the conditional probability density function (PDF) of $\boldsymbol{y}$ given $s$ and $x$ is written as
\begin{equation}\label{eq:condPDF_s_x}
     f(\boldsymbol{y}|x ,s) =  \frac{1}{\pi ^{N_r}} \exp\left\{ -|\boldsymbol{y} - s\boldsymbol{g}(x) |^2 \right\} .
\end{equation}
One method to obtain the maximum likelihood (ML) estimate is the expectation maximization algorithm where the BD signal $x$ is a latent variable. The ML estimate is given by 
\begin{equation*}
    \hat{s}^{\scalebox{0.7}{ML}} = \frac{\boldsymbol{g}^H(x) \boldsymbol{y}}{\|\boldsymbol{g}(x)\|^2} ,
\end{equation*}
and its estimation error is
\begin{equation*}
    \varepsilon = \hat{s}^{\scalebox{0.7}{ML}} - s= \frac{\boldsymbol{g}^H(x) \boldsymbol{n}}{\|\boldsymbol{g}(x)\|^2}.
\end{equation*}
with its variance represented as
\begin{equation*}
    \sigma_{\varepsilon}^2 = \mathbb{E} \{\varepsilon \varepsilon^H\} = \frac{1}{\|\boldsymbol{g}(x)\|^2}.
\end{equation*}
Substituting the ML estimate for the unknown ambient signal $s$ into Eq.~\eqref{eq:received-signal:y} gives
\begin{equation*} \label{eq:received-signal:y_estimateS}
\begin{aligned} 
    \boldsymbol{y} &= (\hat{s}^{\scalebox{0.7}{ML}} -  \varepsilon)\boldsymbol{g}(x)  + \boldsymbol{n} \\
    &=\hat{s}^{\scalebox{0.7}{ML}} \boldsymbol{g}(x) + \Tilde{\boldsymbol{n}} ,
\end{aligned}
\end{equation*}
where the noise term, written as
\begin{equation*}
    \Tilde{\boldsymbol{n}} \triangleq  \boldsymbol{n} - \varepsilon\boldsymbol{g}(x)  =
    \left( \boldsymbol{I} - \frac{\boldsymbol{g}(x) \boldsymbol{g}(x)^H}{\|\boldsymbol{g}(x)\|^2}\right) \boldsymbol{n},
\end{equation*}
follows multivariate complex Gaussian distribution with mean and variance given by
\begin{subequations}
\begin{gather*}
    \mathbb{E}\{\Tilde{\boldsymbol{n}}\} = \left( \boldsymbol{I} - \frac{\boldsymbol{g}(x) \boldsymbol{g}(x)^H}{\|\boldsymbol{g}(x)\|^2}\right) \mathbb{E}\{\boldsymbol{n}\} = \boldsymbol{0} ,\\
         \mathbb{E}\{\Tilde{\boldsymbol{n}} \Tilde{\boldsymbol{n}}^H\} 
          =  \boldsymbol{I} - \frac{\boldsymbol{g}(x) \boldsymbol{g}(x)^H}{\|\boldsymbol{g}(x)\|^2} \triangleq \boldsymbol{G}(x).
\end{gather*}
\end{subequations}
As we can observe, the matrix $\boldsymbol{G}(x)$ is a singular idempotent matrix of rank $N_r-1$.
Hence, $\Tilde{\boldsymbol{n}}$ follows degenerate multivariate complex Gaussian distribution. 
In such case, the conditional PDF in Eq.~\eqref{eq:condPDF_s_x} is defined in the $(N_r-1)$-dimensional affine subspace where the Gaussian distribution can support, which is given by
\begin{equation*}
    \begin{aligned}
        &f(\boldsymbol{y}|x,s=\hat{s}^{\scalebox{0.7}{ML}}) \\ \equiv &\frac{\exp\left\{- \left(\boldsymbol{y} - \boldsymbol{g}(x) \hat{s}^{\scalebox{0.7}{ML}}\right)^H \boldsymbol{G}(x)^{\dagger}  \left(\boldsymbol{y} - \boldsymbol{g}(x)\hat{s}^{\scalebox{0.7}{ML}}\right)  \right\}}{\pi^{N_r} \mathrm{det}^*[\boldsymbol{G}(x)]} \\
        =&   \pi^{-N_r} \exp\left\{- \boldsymbol{y}^H \boldsymbol{G}(x)\boldsymbol{y}  \right\}   ,
        \end{aligned}
\end{equation*}
where $\mathrm{det}^*[\boldsymbol{G}(x)]$ is the pseudo-determinant which is equal to 1~\cite{minka1998inferring}, and $\boldsymbol{G}(x)^{\dagger}$ denotes its Moore–Penrose generalized inverse which in our case is $\boldsymbol{G}(x)$ itself~\cite[sec.~7.3]{Horn}.
Then, taking logarithm of the PDF gives 
\begin{equation*} 
    \begin{aligned}
        &\ln{f(\boldsymbol{y}|x,s=\hat{s}^{\scalebox{0.7}{ML}})} \\=& - \boldsymbol{y}^H \boldsymbol{G}(x) \boldsymbol{y} - N_r \ln \pi .
    \end{aligned}
\end{equation*}

\section{Proof of Proposition II} \label{appendix:matrixM}
The matrix $\boldsymbol{M}$ is spanned by vectors $\boldsymbol{g}_0$ and $\boldsymbol{g}_1$ so that it is a rank-2 matrix. We assume the eigenvalue value decomposition of matrix $\boldsymbol{M}$ is given by
\begin{equation*}
    \boldsymbol{M} = \boldsymbol{U} \boldsymbol{\Lambda} \boldsymbol{U}^H,
\end{equation*}
where $\boldsymbol{\Lambda}$ is a diagonal matrix with the eigenvalues $[\kappa_1, \kappa_2, \boldsymbol{0}_{1\times(N_r-2)}]$ as its diagonal elements and $\boldsymbol{U}$ is an unitary matrix formed by corresponding eigenvectors of $\boldsymbol{M}$. We further denote two eigenvectors corresponding to $\kappa_1$ and $\kappa_2$ as $\boldsymbol{u}_1$ and $\boldsymbol{u}_2$ for later use, respectively.
For obtaining two non-zero eigenvalues, we invoke the theorem that the non-zero eigenvalues of matrices $\boldsymbol{AB}$ are the same as these of matrices $\boldsymbol{BA}$ \cite[Theorem 1.3.22]{Horn}.  To this end, we rewrite
\begin{equation*}
    \begin{aligned}
        \boldsymbol{M} &= \frac{\boldsymbol{g}_0^{} \boldsymbol{g}_0^H }{\|\boldsymbol{g}_0^{}\|^2} - \frac{\boldsymbol{g}_1^{} \boldsymbol{g}_1^H }{\|\boldsymbol{g}_1^{}\|^2} \\ 
        &= \left[\begin{array}{c;{2pt/2pt}c}
   \frac{\boldsymbol{g}_0^{}}{\|\boldsymbol{g}_0^{}\|^2}   & -\frac{\boldsymbol{g}_1^{}}{\|\boldsymbol{g}_1^{}\|^2} \\
   \end{array} \right]
   \left[\begin{array}{c}  \boldsymbol{g}_0^H \\[4pt]    \boldsymbol{g}_1^H \end{array}  \right] .
    \end{aligned}
\end{equation*}
It has the same non-zero eigenvalues as the $2\times2$ matrix written as
\begin{equation*}
   \left[\begin{array}{c}  \boldsymbol{g}_0^H \\[4pt]    \boldsymbol{g}_1^H \end{array}  \right] \left[\begin{array}{c;{2pt/2pt}c}
   \frac{\boldsymbol{g}_0^{}}{\|\boldsymbol{g}_0^{}\|^2}   & -\frac{\boldsymbol{g}_1^{}}{\|\boldsymbol{g}_1^{}\|^2} \\
   \end{array} \right]
   = \left[\begin{array}{c;{2pt/2pt}c}  1  & -\frac{\boldsymbol{g}_0^H \boldsymbol{g}_1^{}}{\|\boldsymbol{g}_1\|^2} \\[6pt]\hdashline[2pt/2pt]
   \frac{\boldsymbol{g}_1^H \boldsymbol{g}_0^{}}{\|\boldsymbol{g}_0\|^2}  &-1
   \end{array} \right] .
\end{equation*}
Thus, two eigenvalues can be readily given by
\begin{equation*}
   \kappa_1 =  \kappa = \sqrt{1 - \frac{|\boldsymbol{g}_0^H \boldsymbol{g}_1^{}|^2}{\|\boldsymbol{g}_1^{}\|^2 \|\boldsymbol{g}_1^{}\|^2}}, \quad \mathrm{and} \quad \kappa_2 = -\kappa ,
\end{equation*}
with their corresponding eigenvectors represented in Eq.~\eqref{eq:eigenvectorM}
.

\ifCLASSOPTIONcaptionsoff
  \newpage
\fi

\bibliographystyle{IEEEtran}

\bibliography{AmBC}

\end{document}